\documentclass[prd, showpacs, twocolumn,superscriptaddress]{revtex4}

\usepackage{graphicx}

\def\lsim{\raise0.3ex\hbox{$\;<$\kern-0.75em\raise-1.1ex
\hbox{$\sim\;$}}}
\def\gsim{\raise0.3ex\hbox{$\;>$\kern-0.75em\raise-1.1ex
\hbox{$\sim\;$}}}

\begin{document}
\title{
Testing for Large Extra Dimensions with Neutrino Oscillations
}
\author{P.~A.~N.~Machado}
\email{accioly@fma.if.usp.br}
\affiliation{
Instituto de F\'{\i}sica, Universidade de S\~ao Paulo, 
 C.\ P.\ 66.318, 05315-970 S\~ao Paulo, Brazil}
\author{H.~Nunokawa}
\email{nunokawa@puc-rio.br} \affiliation{Departamento de F\'{\i}sica,
  Pontif{\'\i}cia Universidade Cat{\'o}lica do Rio de Janeiro,
  C. P. 38071, 22452-970, Rio de Janeiro, Brazil}
\author{R.~Zukanovich Funchal} \email{zukanov@if.usp.br} \affiliation{
  Instituto de F\'{\i}sica, Universidade de S\~ao Paulo,
  C.\ P.\ 66.318, 05315-970 S\~ao Paulo, Brazil}
\vglue 1.4cm

\begin{abstract} 
We consider a model where sterile neutrinos can propagate in a large
compactified extra dimension giving rise to Kaluza-Klein (KK) modes
and the standard model left-handed neutrinos are confined to a
4-dimensional spacetime brane.  The KK modes mix with the standard
neutrinos modifying their oscillation pattern.  We examine former and
current experiments such as CHOOZ, KamLAND, and MINOS to estimate the
impact of the possible presence of such KK modes on the determination
of the neutrino oscillation parameters and simultaneously obtain
limits on the size of the largest extra dimension.  We found that the
presence of the KK modes does not essentially improve the quality of the
fit compared to the case of the standard oscillation.  By combining
the results from CHOOZ, KamLAND and MINOS, in the limit of a vanishing
lightest neutrino mass, we obtain the stronger bound on the size of
the extra dimension as $\sim 1.0(0.6)$ $\mu$m at 99\% C.\ L.\ for normal
(inverted) mass hierarchy.  If the lightest neutrino mass turn out to 
be larger, 0.2 eV, for example, we obtain the bound $\sim 0.1$ $\mu$m.
We also discuss the expected sensitivities on the size of the extra
dimension for future experiments such as Double CHOOZ, T2K and
NO$\nu$A.
\end{abstract} 

\maketitle

\section{Introduction}
\label{sec:intro}

Our observable 1+3-dimensional universe could be a surface, the brane,
embedded in a dimensionally richer 1+3+$d$-dimensional spacetime
($d$ being the number of extra dimension), the bulk. 
This intriguing idea can be motivated by string theory, where at
least 6 extra spatial dimensions are required, in particular, by
stringy inspired models designed to address the disparity between the
electroweak ($\sim 1$ TeV) and the gravity ($\sim 10^{16}$ TeV)
scales.  There are two basic scenarios commonly evoked to generate the
hierarchy between these two fundamental scales of nature: either by
suggesting the source of the hierarchy to be the volume of a flat
extra dimensional space~\cite{ADD} or the strong curvature of that
space~\cite{RS}.

In this paper we are interested in constraining the large extra dimension
(LED) scenario~\cite{ADD} in connection with neutrino physics since
right handed neutrinos (standard model (SM) singlet fields) in this case
can, as well as gravity, propagate in the bulk.  Tabletop
experiments devised to test for deviations of Newtonian gravity can
only probe LED up to submillimeter sizes. The most stringent upper limit
given by a torsion pendulum instrument is 200 $\mu$m at 95\% C.\ L.\  for
the size of the largest flat extra dimension regardless of
the number of $d$~\cite{Hoyle}. Neutrino physics can be considerably 
more sensitive to LED. 

We should, however, mention that astrophysical bounds on 
LED are in general much more stringent 
(see e.g., \cite{Hannestad:2003yd}, and references therein)
than the ones obtained by the terrestrial experiments 
including that from collides. 
However, these astrophysical bounds are not completely 
model independent and therefore, we believe that 
studying the possible impact of LED which can be 
probed (independently from astrophysical constraints) 
by terrestrial experiments is still worthwhile.

There are mounting evidences from several solar~\cite{solar},
atmospheric~\cite{atmos} and
terrestrial~\cite{k2k,kl03,kl11,minos06,minos08,MINOS:Nu2010} neutrino
experiments that neutrinos undergo flavor oscillations due to mass and
mixing.  As it was shown
in~\cite{Dienes:1998sb,ArkaniHamed:1998vp,Dvali:1999cn,Mohapatra-et-al,Barbieri-et-al},
LED can have strong impact on neutrino oscillation probabilities.
However, since the current neutrino data mentioned above are perfectly
consistent with the standard three flavor oscillation scheme, the
effect of LED, if it exists, is expected to be present only as a
subdominant effect on top of the usual oscillation.  Therefore, as was
done in ~\cite{Davoudiasl:2002fq}, in this work we assume that LED
effect would only perturb somewhat the standard oscillation pattern
and try to constrain LED using the current oscillation data.

In this paper, we studied the possible impact of LED on the former and
current oscillation experiments CHOOZ~\cite{Apollonio:2002gd},
KamLAND~\cite{kl03,kl11} and MINOS~\cite{minos06,minos08,MINOS:Nu2010}
in order to obtain the upper bound on the size of the largest extra
dimension, which turns out to be submicrometer range.  We do not
consider solar and atmospheric neutrino data in this work because the
analysis would become much more complicated due to the matter effect
and also because we expect similar bounds from these data (see
Sec.~\ref{sec:conclusions}).  We also calculate the expected
sensitivities on LED for future experiments such as Double
CHOOZ~\cite{Ardellier:2004ui}, T2K~\cite{T2K}, and
NO$\nu$A~\cite{Ayres:2004js,Yang_2004}.

This paper is organized as follows. In Sec.~\ref{sec:form} we
describe the framework of our study of neutrino oscillations with 
LED. In Sec.~\ref{sec:current} and \ref{sec:future} we discuss the 
best current and future limits that can be established on the size of 
the largest extra dimension from neutrino oscillation data.
Finally Sec.~\ref{sec:conclusions} is devoted to discussions and 
general conclusions.
In Appendix~\ref{appendix-solution} we describe the solution 
of the neutrino evolution equation for a constant matter potential
whereas in Appendix~\ref{appendix-simulations}
we describe the details of our $\chi^2$ analysis. 

\section{Neutrino Oscillation Formalism with LED}
\label{sec:form}

We consider here the model discussed in 
Refs.~\cite{Barbieri-et-al,Mohapatra-et-al,Davoudiasl:2002fq} 
where the 3 standard model  active
left-handed neutrinos fields $\nu^{(0)}_{\alpha L}$
($\alpha=e,\mu,\tau$), as well as all the other SM fields, including
the Higgs, are confined to propagate in a 4-dimensional brane, while 3
families of SM singlet fermion fields can propagate in a higher
dimensional bulk, with at least two compactified extra dimensions ($d
\ge 2$). We will assume that one of these extra dimensions is compactified on a
circle of radius $a$, much larger than the size of the others so that
we can in practice use a 5-dimensional treatment.

By this assumption, our bounds are always more conservative
than the ones obtained by assuming all the LED radius $a$ are
the same for $d \ge 2$. In other words, if we have adopted
the same assumption (of equal raduis for all LED),
we should have obtained stronger bounds on $a$ for $d \ge 2$
because the conversion into KK modes would be more efficient
under such an assumption for $d \ge 2$.

The 3 bulk fermions will have Yukawa couplings with the SM Higgs and
the brane neutrinos ultimately leading to Dirac masses and mixings
among active species and sterile Kaluza-Klein (KK) modes.  
 The 4-dimensional Lagrangian which describes the charged current (CC)
interaction of the brane neutrinos with the $W$ as well as the
mass term resulting from these couplings with the bulk fermions in the
brane, after electroweak symmetry breaking and dimensional reduction,
can be written as~\cite{Davoudiasl:2002fq}

\begin{widetext}
\begin{eqnarray}
\hskip -0.5cm
\mathcal{L_{\text{eff}}} \, &= &\, \mathcal{L_{\text{mass}}} +
\mathcal{L_{\text{CC}}} \nonumber \\ &= & \displaystyle
\sum_{\alpha,\beta}m_{\alpha\beta}^{D}\left[\overline{\nu}_{\alpha
    L}^{\left(0\right)}\,\nu_{\beta R}^{\left(0\right)}+\sqrt{2}\,
  \sum_{N=1}^{\infty}\overline{\nu}_{\alpha
    L}^{\left(0\right)}\,\nu_{\beta
    R}^{\left(N\right)}\right]  
 +  \sum_{\alpha}\sum_{N=1}^{\infty}\displaystyle
\frac{N}{a}\, \overline{\nu}_{\alpha L}^{\left(N\right)} \,
\nu_{\alpha R}^{\left(N\right)} 
\,+  \displaystyle
\frac{g}{\sqrt{2}}\,\sum_{\alpha} \,
\overline{l_\alpha}\gamma^{\mu}\left(1-\gamma_{5}\right)\nu_{\alpha}^{\left(0\right)}\,
W_{\mu}+\mbox{h.c.},
\end{eqnarray}
\end{widetext}
where the Greek indices $\alpha,\beta = e,\mu,\tau$, the capital Roman
index $N=1,2,3,...,\infty$, $m_{\alpha \beta}^{D}$ is a Dirac mass
matrix, $\nu^{(0)}_{\alpha R}$, $\nu^{(N)}_{\alpha R}$ and
$\nu^{(N)}_{\alpha L}$ are the linear combinations of the bulk fermion
fields that couple to the SM neutrinos $\nu^{(0)}_{\alpha L}$.

After performing unitary transformations in order to diagonalize
$m^{D}_{\alpha \beta}$ we arrive at the neutrino evolution equation
(\ref{eq:evol}) that can be solved to obtain the eigenvalues
$\lambda_j^{(N)}$ and amplitudes $W_{ij}^{(0N)}$ (see
Appendix~\ref{appendix-solution}), so that the transition probability
of $\nu_{\alpha}^{(0)}$ into $\nu_{\beta}^{(0)}$ (subscript $L$ is
omitted) at a distance $L$ from production,

\begin{equation}
P(\nu_\alpha^{(0)} \to \nu_\beta^{(0)};L) = \vert {\cal{A}}_{\nu_\alpha^{(0)}\to 
\nu_\beta^{(0)}} (L)\vert^2 \, ,
\end{equation}
can be given in terms of the transition amplitude
\begin{eqnarray}
{\cal{A}}_{\nu_\alpha^{(0)}\to \nu_\beta^{(0)}} (L) & = & \displaystyle
\sum_{i,j,k=1}^{3}\sum_{N=0}^{\infty} U_{\alpha i} U_{\beta k}^{*}
W_{ij}^{(0N)*}W_{kj}^{(0N)} \nonumber \\
& & \times \exp \left(i\frac{\lambda_j^{(N)2}L}{2Ea^2} \right)\, ,
\label{eq:amplitude}
\end{eqnarray}
where $E$ is the neutrino energy, $L$ is the baseline distance, 
$\lambda_j^{(N)}$ is the eigenvalues of the Hamiltonian 
of the evolution Eq.~(\ref{eq:evolution}) in the Appendix, and  
$U$ and $W$ are the mixing matrices for active and KK neutrino modes,
respectively.

This transition probabilities, even in vacuum, depend on the neutrino
mass hierarchy since both $W^{(0N)}_{ij}$ and $\lambda^{(N)}_j$ are
functions of the dimensionless parameter 
$\xi_j \equiv \sqrt{2} \, m_j \,a$, where $m_j$ $(j=1,2,3)$ are 
the neutrino masses. We will consider here two possibilities for the mass
hierarchy: normal hierarchy (NH) with $m_3 > m_2 > m_1=m_0$ and
inverted hierarchy (IH) with $m_2> m_1 > m_3=m_0$. As $m_0$ increases
NH and IH become degenerate. We define the mass squared differences as
$\Delta m^2_{ij} \equiv m^2_i - m^2_j$ ($i,j = 1,2,3$).

\begin{figure}
\begin{center}
\includegraphics[width=0.47\textwidth]{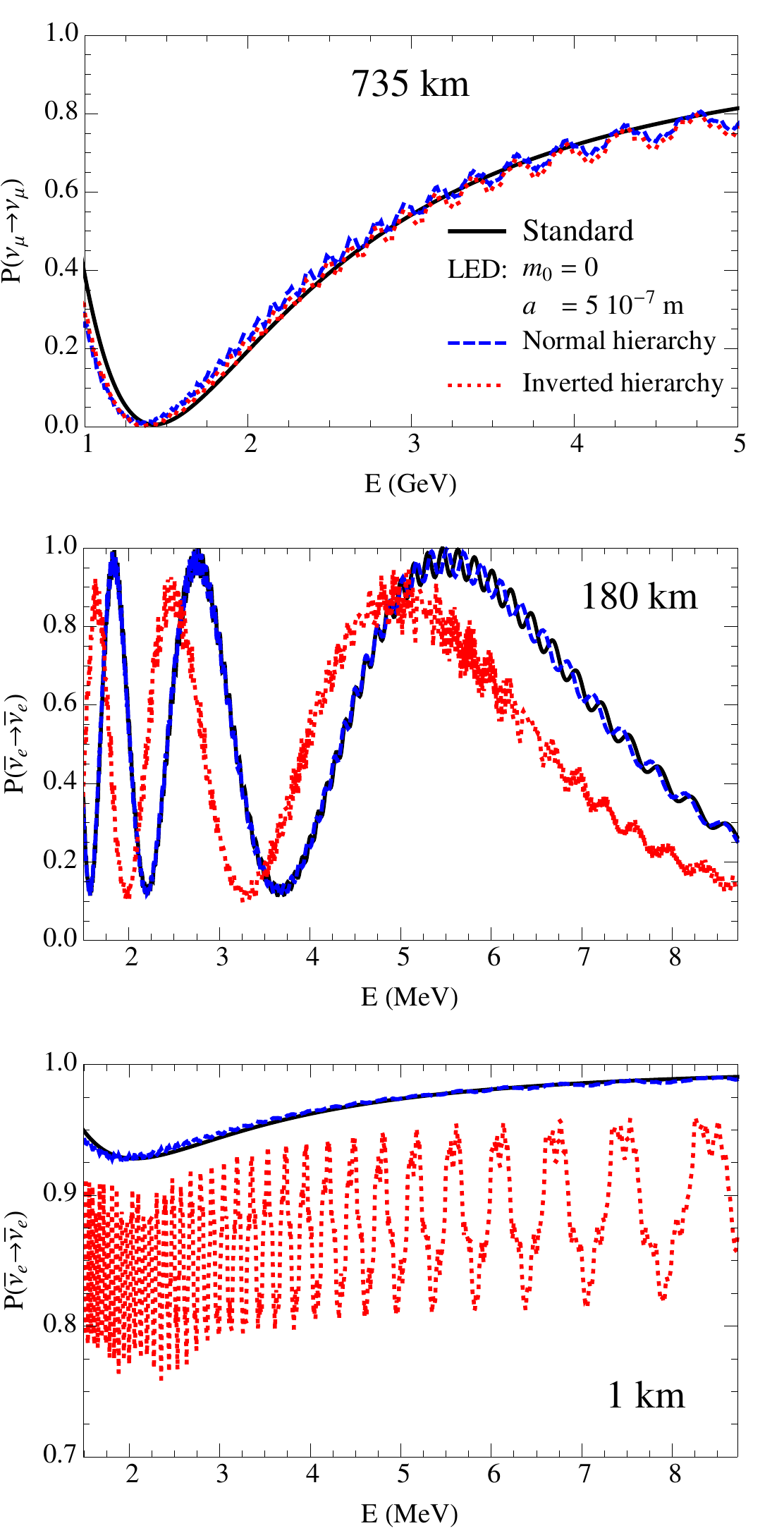}
\end{center}
\vglue -0.6cm
\caption{In the top (middle and bottom) panel we show the survival
  probability for $\nu_\mu$ ($\bar \nu_e$) as a function of the
  neutrino energy $E$ for the baseline  $L = 735$ km ($180$ km and $1$ km) 
for $a=0$ (no  LED, black curve) and $a=0.5$ $\mu$m for NH (blue dashed
  curve) and IH (red dotted curve).  The other oscillation parameters
  were set to $\sin^2 \theta_{12}=0.32$, $\sin^2 \theta_{23}=0.5$,
  $\sin^2 2\theta_{13}=0.07$, $\Delta m^2_{21}=7.59 \times
  10^{-5}$ eV$^2$, and $\vert \Delta m^2_{31} \vert=2.46 \times
  10^{-3}$ eV$^2$.
The lightest neutrino mass, $m_0$, was set to zero. }
\label{fig:prob}
\end{figure}

To understand qualitatively the results to be presented in
Secs.~\ref{sec:current} and \ref{sec:future} we discuss here what
is to be expected of the effects of LED on the survival probabilities.
In Fig.~\ref{fig:prob} we show the survival probabilities for
$\nu_\mu$ and $\bar \nu_e$ as a function of the neutrino energy $E$ in
vacuum for NH and IH for MINOS ($735$ km), KamLAND ($180$ km) and
CHOOZ ($1$ km), assuming $m_0=0$ and $a= 0.5$ $\mu$m. There are three
basic effects of LED: a displacement of the minima with respect to the
standard survival probabilities, a global reduction of the flavor
survival probabilities as SM neutrinos can oscillate into KK modes and
the appearance of extra wiggles on the probability pattern due to the
fast oscillations to these new massive modes.

When matter effects can be ignored, the impact of LED in the survival
amplitude, to leading order in $\xi_i$, is such that ${\cal{A}}^{\rm
  (LED)}_{\nu_\alpha^{(0)}\to \nu_\alpha^{(0)}} \propto \sum_i \xi_i^2
\, \vert U_{\alpha i}\vert^2 $ (see Eq.~\ref{eq:vac}). 
Therefore, roughly speaking, 
in order to modify the standard probability by say $\sim$ 10\% by 
the effect due to LED, at least one of the $\xi_i^2$ should be 
order of $\sim$ 0.1. 
Since we know, from atmospheric neutrino oscillation, 
that at least one neutrino has a mass $m_i \sim 0.05$ eV, 
we can estimate that $\xi_i^2 \sim 0.1$ implies 
$a \sim 5 \, \rm eV^{-1} = 1$ $\mu$m. 
So one can expect the terrestrial experiments to be sensitive around
this scale, which is consistent with our results discussed in the next
section.

From now on we will drop the $(0)$ superscript when referring to 
flavor oscillations.
In the case of $\nu_\mu \to \nu_\mu$, from Fig.~\ref{fig:prob} we see
that the effect of LED is basically the same for NH and IH. This is
because  ${\cal{A}}^{\rm  (LED)}_{\nu_\mu^{(0)}\to \nu_\mu^{(0)}} \propto \sum_i \xi_i^2
\, \vert U_{\mu i}\vert^2 $ is of the same order for NH 
(mainly driven by $\xi_3$) and IH (mainly driven by $\xi_2$). 
On the other hand, in the case of $\bar\nu_e \to \bar \nu_e$ the
effect of LED is significantly larger for IH than NH 
since 
${\cal{A}}^{\rm  (LED)}_{\bar \nu_e^{(0)}\to \bar \nu_e^{(0)}} \propto \sum_i \xi_i^2
\, \vert U_{e i}\vert^2 = \sum_{i=1,2} \xi_i^2 \,\vert U_{e i}\vert^2 + \xi^2_{3} \, \sin^2 \theta_{13}$ 
is suppressed due to small $\sin^2 \theta_{13}$ 
for NH (since $\xi_3 \gg \xi_1,\xi_2$ for vanishing $m_0$) 
whereas for IH the dominant LED contributions 
(due to $\xi_1$ and $\xi_2$) are not suppressed. 

In this paper we do not consider the appearance channels 
such as $\nu_\mu \to \nu_e$ and $\bar{\nu}_\mu \to \bar{\nu}_e$
due to the following reasons. 
First of all, when $\theta_{13}$ is zero or much smaller than the current bound, 
we found that the impact of LED for these appearance channels is very small 
compared with that of the disappearance modes considered in this work.
In principle, even if $\theta_{13}$ is zero, 
LED can induce a flavor transition such as 
$\nu_\mu \to \nu_e$ (for T2K and NO$\nu$A) 
through the right handed KK modes 
but such a transition is a kind of second order effect 
(this is because, ignoring oscillation driven by solar parameters, 
$\nu_\mu$ would not be converted directly to $\nu_e$ but only through KK modes) 
whereas the impact of LED for the disappearance channel 
is the first order effect, or it is the consequence 
of the direct transition from active to sterile KK modes. 
This argument applies also to the appearance experiments like
LSND~\cite{Aguilar:2001ty}, Karmen~\cite{Armbruster:2002mp} 
and MiniBOONE~\cite{AguilarArevalo:2009xn}, and 
therefore we do not consider these experiments in this
work, as they do not make any significant 
contribution to improve the bounds we obtained. 

On the other hand, if $\theta_{13}$ is large enough to be 
observed by T2K and NO$\nu$A (in the absence of LED), 
then the impact of LED can be sizable but only as a small
perturbation on top of the standard oscillation
unless we consider LED parameters not allowed by 
the disappearance modes. 
While LED can be potentially harmful in the appearance 
modes for the determination of the mass hierarchy and/or
CP phase delta, we believe that the appearance mode is 
not important (due to much smaller statistics than 
the disappearance ones) in constraining LED.

\section{Current Experimental Limits}
\label{sec:current}

Here we discuss the limits on the size of LED one can obtain from the
former and current neutrino oscillation experiments CHOOZ, KamLAND and
MINOS. We could have considered other terrestrial experiments in our
analysis, but we have restricted ourselves to these three. Regarding
the long baseline experiments, KamLAND and MINOS are currently the
best ones in terms of statistic and systematics.  While the inclusion
of other short baseline experiments could, in principle, improve our
results, we have verified that this improvement is not very
significant, since a large fraction of the uncertainties of these experiments
are correlated.

We do not consider solar and atmospheric neutrino data for simplicity,
and also because we do not expect significant improvement in
constraining LED by adding these data (see
Sec.~\ref{sec:conclusions}).

\subsection{Reactor $\bar \nu_e \to \bar \nu_e$ Experiments: CHOOZ and KamLAND}

The CHOOZ experiment is a former long (for reactor) 
baseline reactor neutrino
oscillation experiment. Its goal was to probe the atmospheric
oscillation parameters in order to shed light on the atmospheric
anomaly~\cite{Apollonio:2002gd}. To achieve that aim, the experiment
detected $\bar\nu_e$ produced by the French CHOOZ nuclear power plant
via the inverse $\beta$-decay reaction $\bar \nu_e +p \to e^{+} +
n$. The $\bar \nu_e$ energy $E$ is estimated from the observed prompt
energy $E_p$ of $e^+$ and nucleon mass difference $M_n - M_p$ as $E
\approx E_p + (M_n - M_p) + \mathcal{O}(E_{\bar\nu_e}/M_n)$, where the
last term corresponds to the neutron recoil. Hence, the reaction has a
1.8 MeV threshold.

The KamLAND (Kamioka Liquid scintillator Anti-Neutrino Detector) is a
reactor neutrino oscillation experiment that operates in the site of
the former Kamiokande experiment in Japan. Since 2003 KamLAND has
observed $\bar \nu_e$ disappearance~\cite{kl03} compatible with the
standard neutrino oscillation scenario, giving strong support to the MSW LMA
solution to the solar neutrino problem reported by the solar neutrino
experiments~\cite{solar}.  
The KamLAND detector observes $\bar \nu_e$ produced by the surrounding
nuclear power reactors via the same inverse $\beta$-decay reaction
described above. 

We have analyzed the last result by CHOOZ~\cite{Apollonio:2002gd} and the
most recent KamLAND data~\cite{kl11}. In fitting CHOOZ (KamLAND) data
we have used the results of the new flux 
calculation for reactor neutrinos~\cite{Mueller:2011nm,reactor-anomaly}
and varied $\Delta m^2_{31}$ and $\theta_{13}$ ($\Delta m^2_{21}$
and $\theta_{12}$) freely. 
We note, however, that the change of the reactor neutrino flux 
to the new one has very little impact on our results in obtaining LED bounds. 
For both experiments, we considered priors on all
other standard oscillation parameters as explained in
Appendix~\ref{appendix-simulations}, except when comparing our
standard KamLAND fit to \cite{kl11}, where we took $\theta_{13} = 0$.

\begin{figure}[h]
\begin{center}
\includegraphics[scale=0.63]{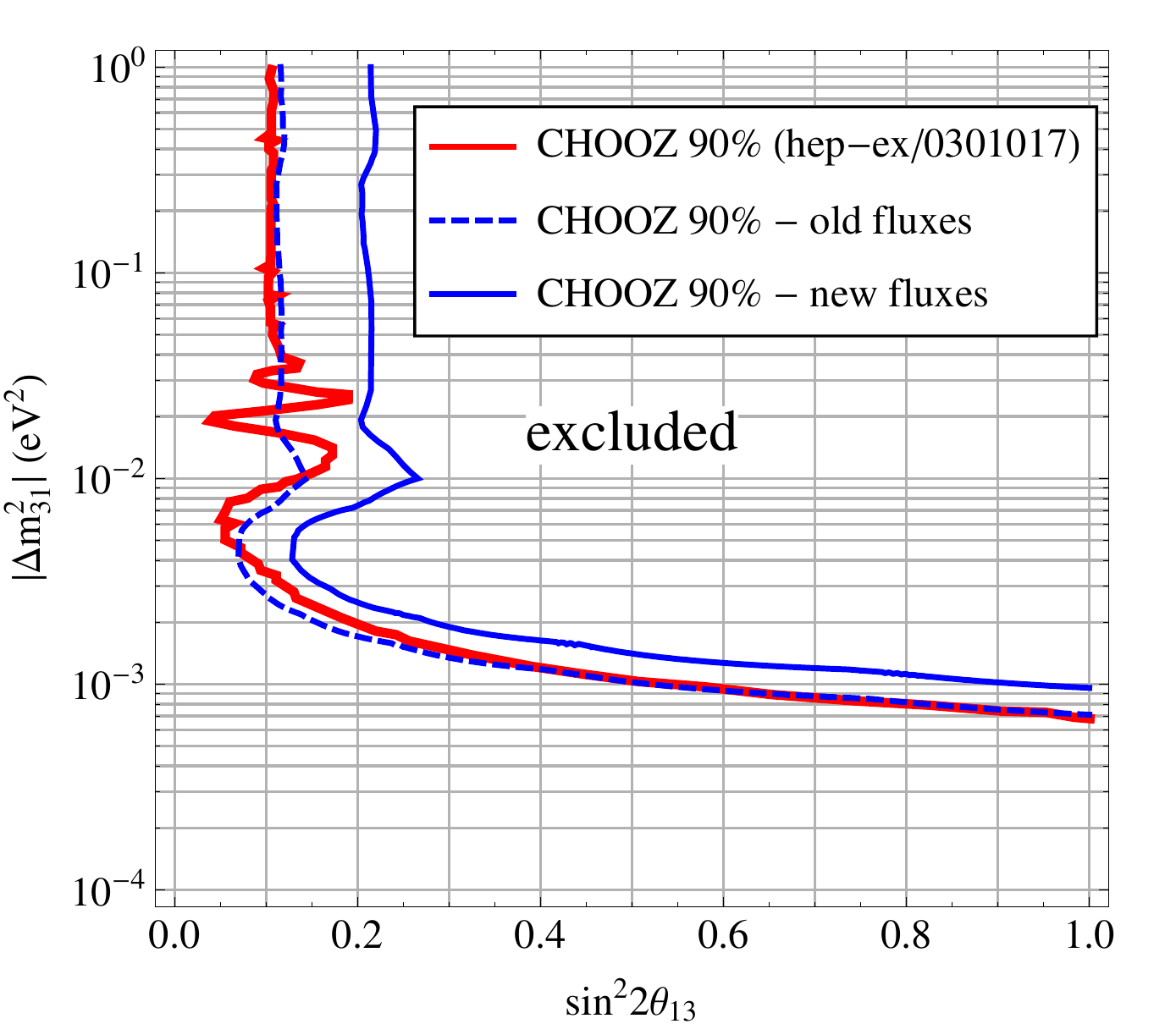}
\end{center}
\vglue -0.6cm
\caption{Allowed regions in the $\sin^{2}2 \theta_{13} - \vert\Delta
  m^2_{31}\vert$ plane obtained by fitting the CHOOZ data at 90\% C.\ L. 
The result of our  simulation using the old reactor fluxes is indicated by 
the dashed blue curve, which is to be compared with 
the result of the analysis A exclusion limit presented
in~\cite{Apollonio:2002gd} (red curve). 
The solid blue curve  represents the allowed region using the updated reactor fluxes
from~\cite{reactor-anomaly}.}
\label{fig:chooz-std}
\end{figure}

In Fig.~\ref{fig:chooz-std} we show the region in the
$\sin^{2}2 \theta_{13} - \vert\Delta m^2_{31}\vert$ plane 
allowed by CHOOZ data at 90\% C.\ L.\ 
for the standard oscillation case with $a=0$.  
While we used the new reactor neutrino 
flux~\cite{Mueller:2011nm,reactor-anomaly} 
to study the impact of LED throughout this paper, 
in order to compare the results of our
analysis with the original results
by the CHOOZ group~\cite{Apollonio:2002gd} (indicated by the solid red curve) 
we show the result obtained by using old flux (dashed blue curve) 
in addition to the one with the new flux (solid blue curve). 
We note that our  simulation using the old reactor fluxes agrees 
reasonably well with that of CHOOZ~\cite{Apollonio:2002gd}. 

We verified that the 
inclusion of LED does not essentially change this region. 
This can be understood if we remember that the main effect of LED is to induce oscillations
to the sterile KK modes. Since CHOOZ basically does not see any
significant deviation of the average $\bar\nu_e\to\bar\nu_e$ probability from
unity and the inclusion of LED can only lower this probability, LED
cannot enlarge the CHOOZ allowed region.

\begin{figure}[h]
\begin{center}
\includegraphics[scale=0.68]{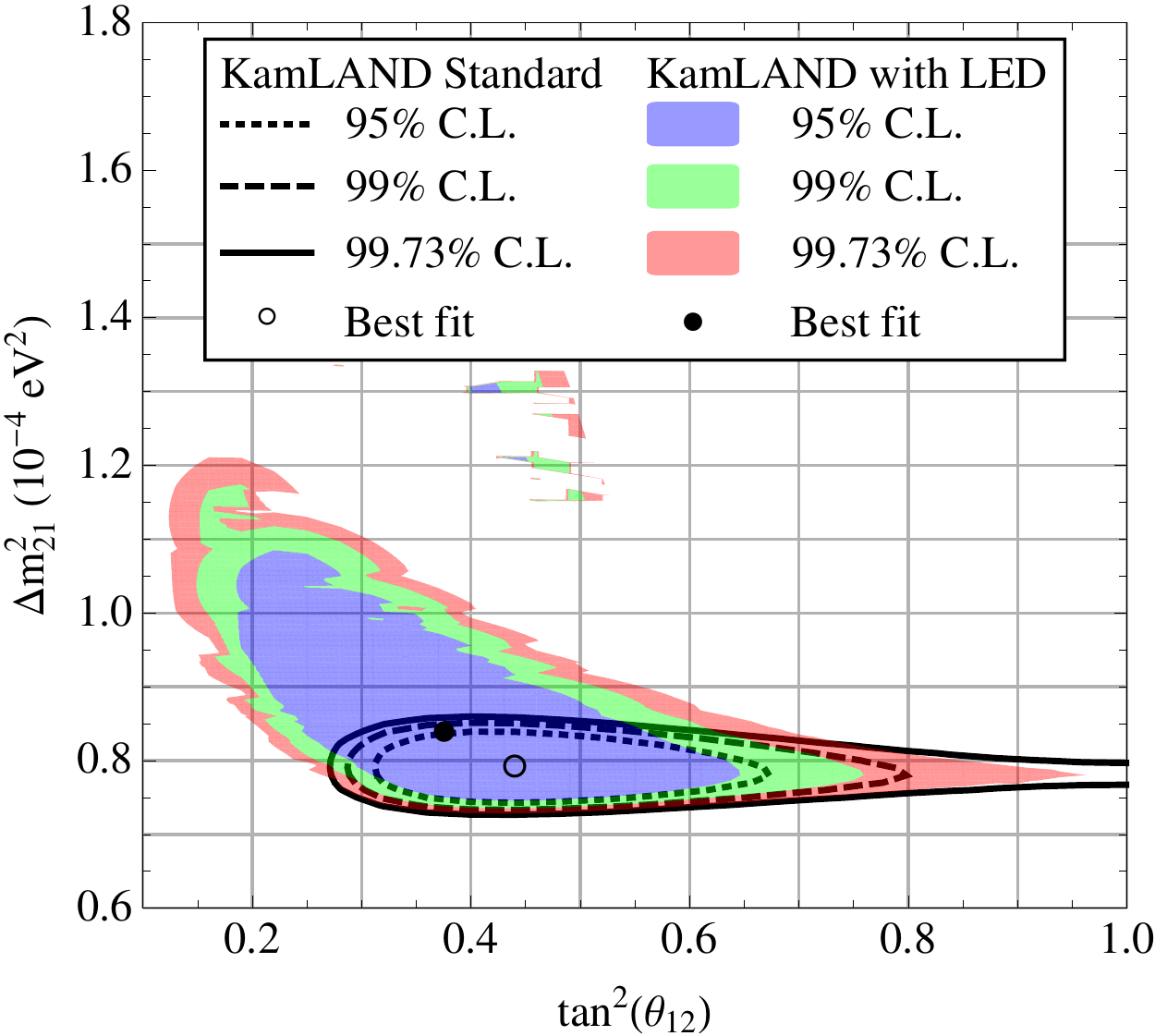}
\end{center}
\vglue -0.6cm
\caption{Allowed regions in the $\tan^{2}\theta_{12} - \Delta
  m^2_{21}$ plane obtained by fitting the KamLAND data at 95\%, 99\% and
  99.73\% C.\ L. We compare the standard oscillation scheme (lines) with
  the LED oscillation scheme (colored regions). }
\label{fig:kl-std}
\end{figure}

In Fig.~\ref{fig:kl-std} we show the regions in 
the $\tan^{2}\theta_{12} - \Delta m^2_{21}$ plane 
allowed by the KamLAND data at 95\%, 99\% and 99.73\%
C.\ L.\  for the standard oscillation case with $a=0$ (indicated by 
the dotted, dashed and solid curves) superimposed on 
the case fitted with LED (shaded colored regions). 
When we fit with LED we have also varied freely $a$, $m_0$ and the mass hierarchy.

We see that our simulation agrees reasonably with the standard result
of Fig. 2 of Ref.~\cite{kl11}, our best fit point corresponds to
$\Delta m^2_{21} =7.84 \times 10^{-5}$ eV$^2$ and $\tan^2\theta_{12}
=0.46$ with $\chi^2_{\rm min}/{\rm dof}= 17.3/15=1.15$.  With LED the
allowed region gets considerably larger, our best fit point here
corresponds to $\Delta m^2_{21} =8.28 \times 10^{-5}$ eV$^2$,
$\tan^2\theta_{12} =0.38$ and $a=0.52$ $\mu$m for NH with $m_0=3.56
\times 10^{-2}$ eV.  However, $\chi^2_{\rm min}/{\rm dof}=
16.8/13=1.29$, so the inclusion of LED does not improve the fit.

\begin{figure}[bhtp]
\begin{center}
\includegraphics[scale=0.62]{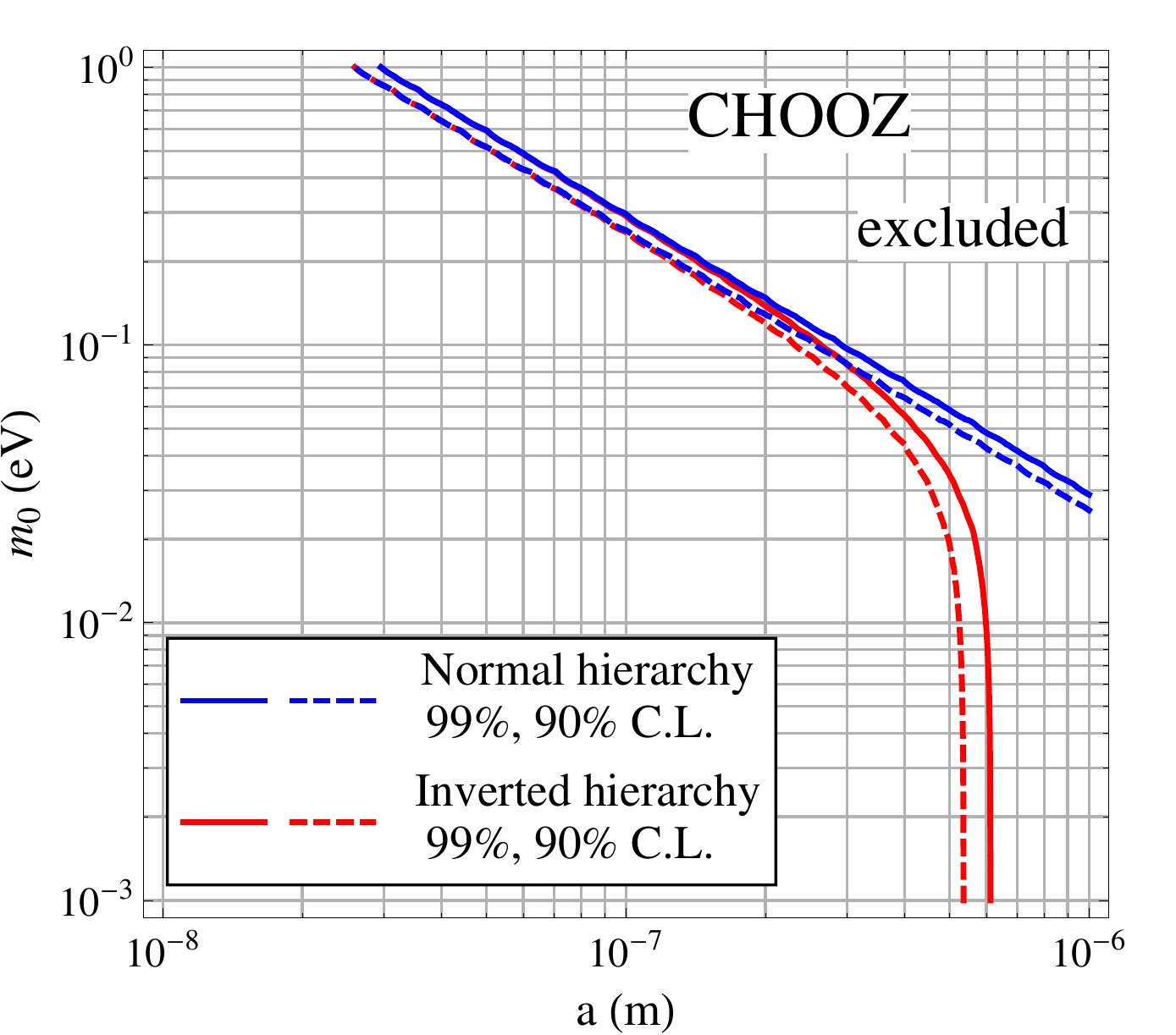}
\end{center}
\vglue -0.6cm
\caption{Excluded regions in the $ a - m_0$ plane  ($m_0$ is the
  lightest neutrino mass) by CHOOZ data at 90\% and 99\% C.\ L.\  for NH
  (blue curves) and IH (red curves). }
\label{fig:chooz-led}
\end{figure}

\begin{figure}[bhtp]
\begin{center}
\includegraphics[scale=0.62]{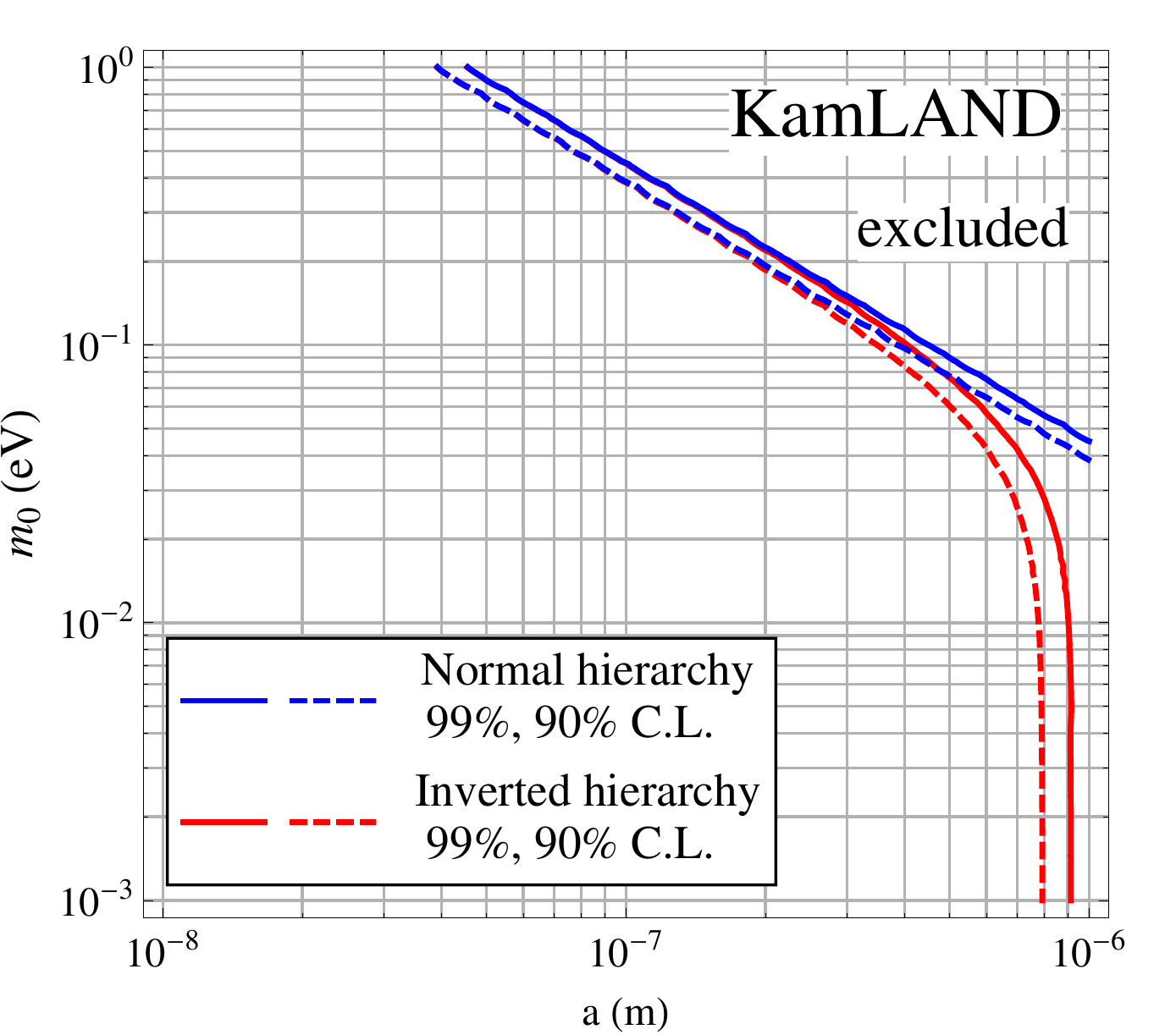}
\end{center}
\vglue -0.6cm
\caption{Same as Fig.~\ref{fig:chooz-led} but excluded by KamLAND data. }
\label{fig:kl-led}
\end{figure}

We have also investigated what region in the $a - m_0$ plane can be
excluded by CHOOZ and KamLAND data. This was calculated for NH and IH
at 90\% (99\%) C.\ L., by imposing $\chi^2>\chi^2_{\rm min} + 4.61 \,(9.21)$,
and is presented in Fig.~\ref{fig:chooz-led} (CHOOZ) and
Fig.~\ref{fig:kl-led} (KamLAND).  As expected the $\bar \nu_e \to \bar
\nu_e$ channel gives a much more stringent limit on LED for the IH
case (see Fig.\ref{fig:prob}). We see that CHOOZ limits are stronger
than KamLAND limits. For some numerical limits see Table~\ref{table:limits}.

We can understand qualitatively the shape of our exclusion curves 
in Figs.~\ref{fig:chooz-led} and \ref{fig:kl-led} as follows. 
If $m_0 \gsim 0.05$ eV, neutrino masses are degenerate and in this case 
the limit has to be proportional to $\xi = \sqrt{2} \, m_0 \, a$, 
i.e., if $\xi> \xi_{\rm max}$ the region is excluded; 
this explains the linear behavior at the upper part of the plots
of Figs.~\ref{fig:chooz-led} and \ref{fig:kl-led}.
If, however, $m_0 << 0.05$ eV, LED will be constrained by $\xi_{2,3}$ (NH) 
or $\xi_{1,2}$ (IH) so the limit will not depend on $m_0$; this explains 
lower part of the plots.

\subsection{Accelerator $\nu_\mu \to \nu_\mu$ Experiment: MINOS}

MINOS (Main Injector Neutrino Oscillation Search) is a neutrino
oscillation experiment at Fermilab that has been running since the 2006
accelerator-beam $\nu_\mu$ disappearance~\cite{minos06,minos08}
supporting the results from K2K~\cite{k2k} and the atmospheric
neutrino experiments~\cite{atmos}. MINOS has a magnetized near
detector with 29 t fiducial mass at 1.04 km from the production target
and a magnetized far detector with a fiducial mass of 4 kt at 735 km.
Recently MINOS has also reported the observation of accelerator-beam
$\bar \nu_\mu$ disappearance~\cite{MINOS:Nu2010}, which we will not
consider in this work due to low statistics.  In MINOS $\nu_\mu$ are
identified by charged current interactions and the sign of the
associated muon produced which is determined by the muon curvature
under the detectors magnetic fields. The main background is due to neutral
current events.

\begin{figure}[bhtp]
\begin{center}
\includegraphics[scale=0.64]{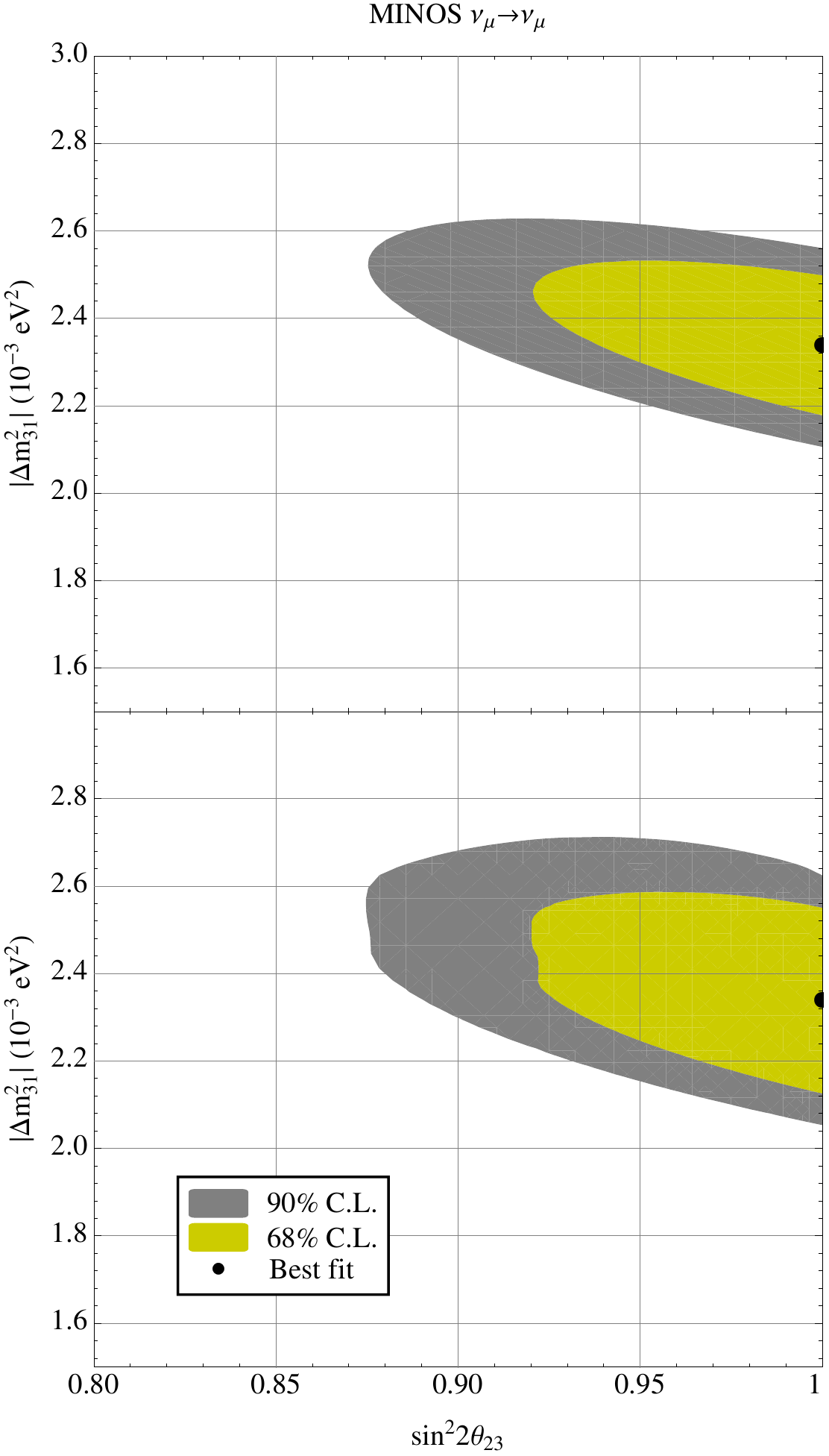}
\end{center}
\vglue -0.7cm
\caption{Allowed region for the standard oscillation parameters in the
$\sin^2 2\theta_{23} - \vert \Delta m^2_{31}\vert$   plane from
  MINOS $\nu_\mu \to \nu_\mu$ data. In the upper panel we assumed no
  LED while in the lower panel we allowed for LED in the fit.}
\label{fig:minos-std}
\end{figure}

We have analyzed the most recent MINOS data in the $\nu_\mu \to
\nu_\mu$ mode~\cite{MINOS:Nu2010}. In Fig.~\ref{fig:minos-std} we show
the allowed regions in the $\sin^2 2\theta_{23} - \vert \Delta
m^2_{31}\vert $ plane at 68\% and 90\% C.\ L. In the upper panel we have
the pure standard oscillation (no large extra dimension allowed,
$a=0$) and in the lower panel we have allowed for LED. In fitting the
data we have varied $\vert\Delta m^2_{31}\vert$ and
$\sin^22\theta_{23}$ freely, and considered priors on all other
standard oscillation parameters (see
Appendix~\ref{appendix-simulations} for further details). When we fit
with LED we have also varied freely $a$, $m_0$ and the mass hierarchy.

Our best fit point in the standard oscillation fit is $\vert \Delta
m^2_{31} \vert =2.39 \times 10^{-3}$ eV$^2$ and $\sin^22\theta_{23}
=1$ with $\chi^2_{\rm min}/{\rm dof}= 12.3/12=1.02$. With LED the
allowed region gets enlarged, however the best fit point remains 
the same with $a=0$, hence any value of $m_0$ is allowed.  The
$\chi^2_{\rm min}/{\rm dof}= 12.3/10=1.23$, so the inclusion of LED
worsens the fit to data.

\begin{figure}[h]
\begin{center}
\includegraphics[scale=0.60]{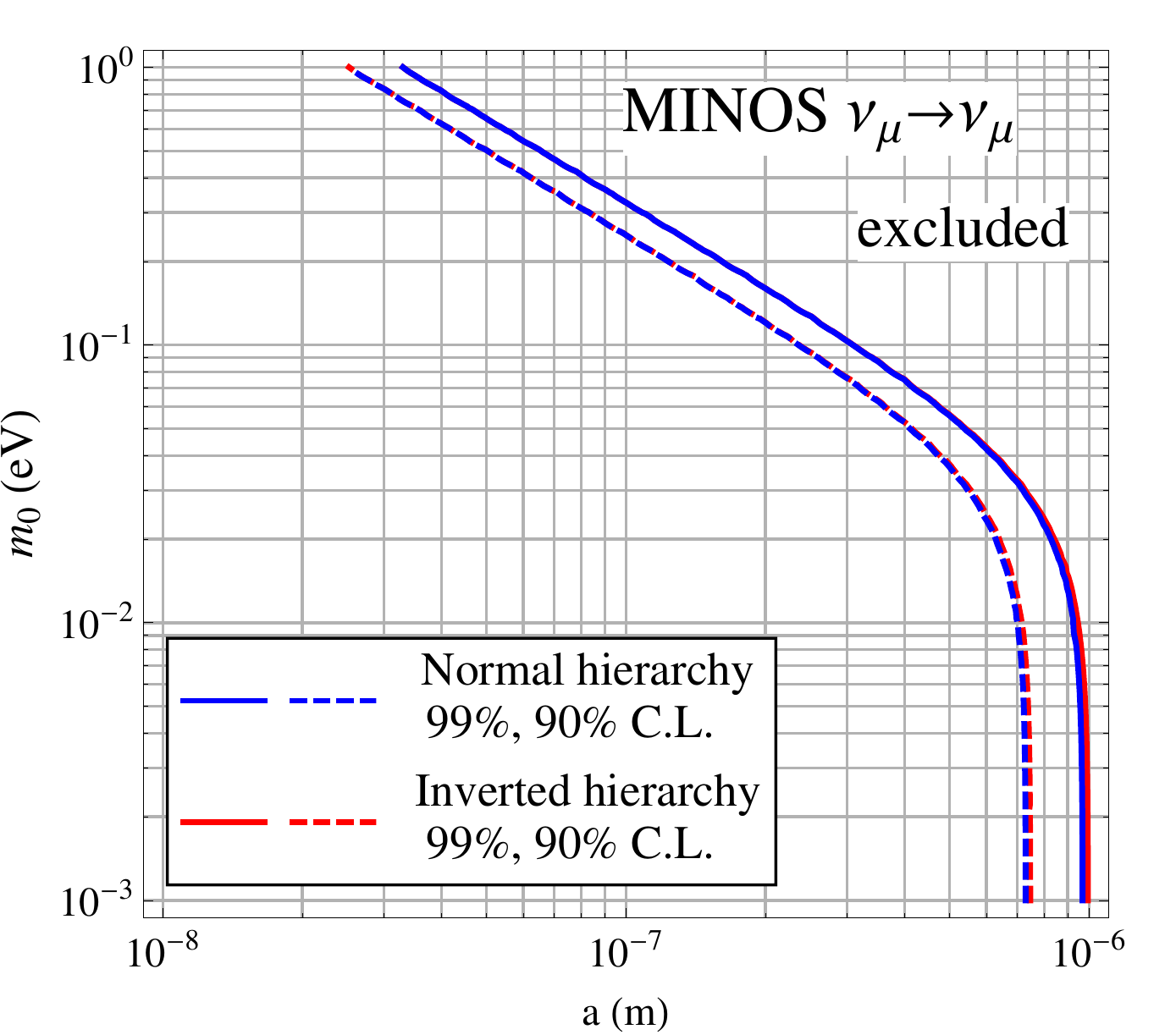}
\end{center}
\vglue -0.6cm
\caption{Same as Fig.~\ref{fig:chooz-led} but excluded by MINOS $\nu_\mu
  \to \nu_\mu$ data.}
\label{fig:minos-led}
\end{figure}

We have investigated what region in the $ a - m_0$ plane can be
excluded by MINOS $\nu_\mu \to \nu_\mu$ data.  In
Fig.~\ref{fig:minos-led} we present the excluded region calculated for 
NH and IH at 90\% and 99\% C.\ L.  As expected the $\nu_\mu \to \nu_\mu$
channel is equally sensitive to NH and IH (see Fig.\ref{fig:prob}).
For some numerical limits, see Table~\ref{table:limits}.

\subsection{CHOOZ, KamLAND and MINOS Combined}
\begin{figure}[bhtp]
\begin{center}
\includegraphics[scale=0.60]{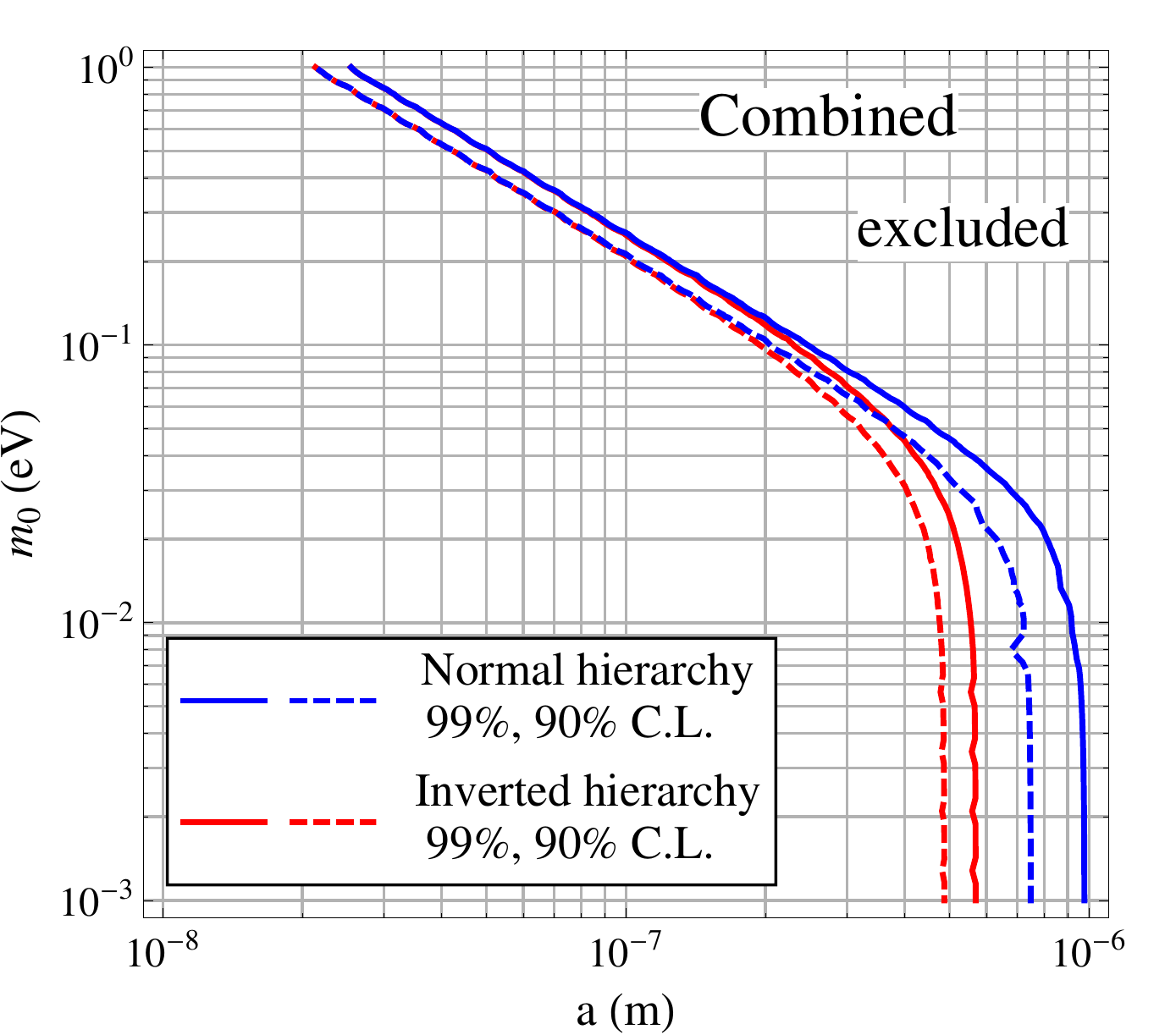}
\end{center}
\vglue -0.6cm
\caption{Same as Fig.~\ref{fig:chooz-led} but excluded by CHOOZ, 
KamLAND and MINOS combined data.}
\label{fig:comb-led}
\end{figure}

We have analyzed MINOS $\nu_\mu\to\nu_\mu$ together with CHOOZ and
KamLAND data by minimizing their added up $\chi^2$ functions letting
all parameters vary freely. The excluded region for LED given by the
combined fit is shown in Fig.~\ref{fig:comb-led}. We see that the
combined fit improves the limits derived until here, except for NH
when $m_0 \to 0$ where the limit is basically that given by MINOS. For
some numerical limits, see Table~\ref{table:limits}.

\section{Future Terrestrial Neutrino Oscillation Experiments}
\label{sec:future}

Here we discuss the possibility of improving the current limits on LED 
by the future neutrino oscillation experiments Double CHOOZ, NO$\nu$A and 
T2K.

\subsection{Reactor $\bar \nu_e \to \bar \nu_e$ Experiment: Double CHOOZ}

The Double CHOOZ experiment~\cite{Ardellier:2004ui}, is a reactor
neutrino oscillation experiment that is being built in France 
which aims to explore the range $0.03<\sin^2 2\theta_{13}<0.2$. There
will be two identical 8.3 t liquid scintillator detectors, one at 400
m and the other at 1.05 km from the nuclear cores.  The expected luminosity
is 400 t GW y. We will consider 3 years of data taking in our
calculations.  In fitting the data we have varied $\vert\Delta
m^2_{31}\vert$ and $\sin^22\theta_{13}$ freely, and considered priors
on all other standard parameters (See Appendix~\ref{appendix-simulations}).

In Fig.~\ref{fig:dc-std} we show our expected sensitivity 
for $\sin^2 2\theta_{13}$ as a function of $\vert \Delta m^2_{31}\vert$
for Double CHOOZ after 3 years for the standard oscillation analysis. 
We have verified that allowing for LED in the fit does not change this 
sensitivity curve as long as $a<0.3$ $\mu$m.  So LED cannot simulate 
a nonzero $\theta_{13}$.

\begin{figure}[bhtp]
\begin{center}
\includegraphics[scale=0.68]{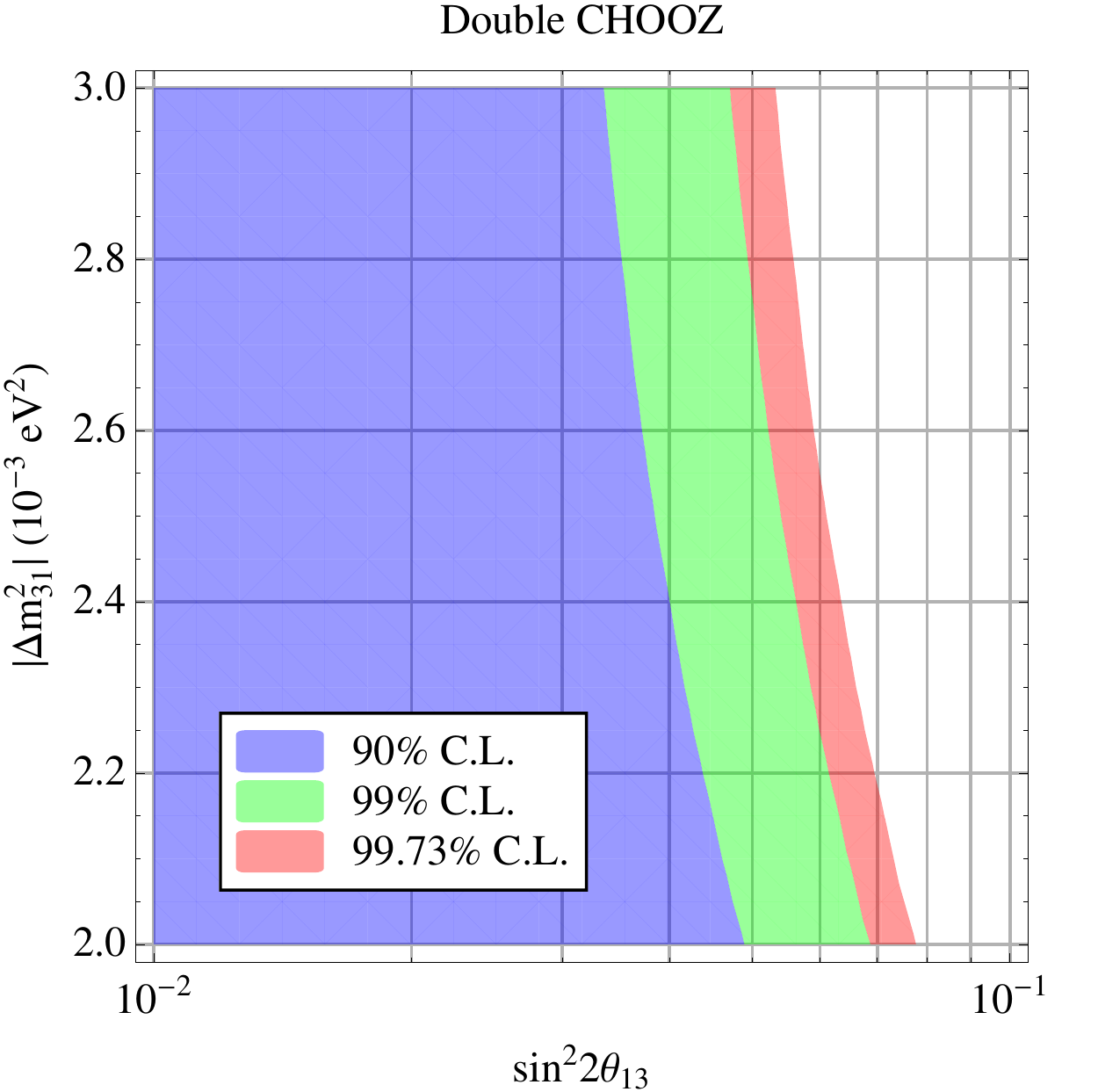}
\end{center}
\vglue -0.6cm
\caption{
Sensitivity to $\sin^2 2\theta_{13}$ predicted for Double
CHOOZ after 3 years, without assuming LED (standard oscillation). 
Here we have assumed as input: $\sin^2 2\theta_{13}=0$, $\vert \Delta m^2_{31}\vert=2.46
\times 10^{-3}$ eV$^2$, and $a=0$.}
\label{fig:dc-std}
\end{figure}

\begin{figure}[bhtp]
\begin{center}
\includegraphics[scale=0.62]{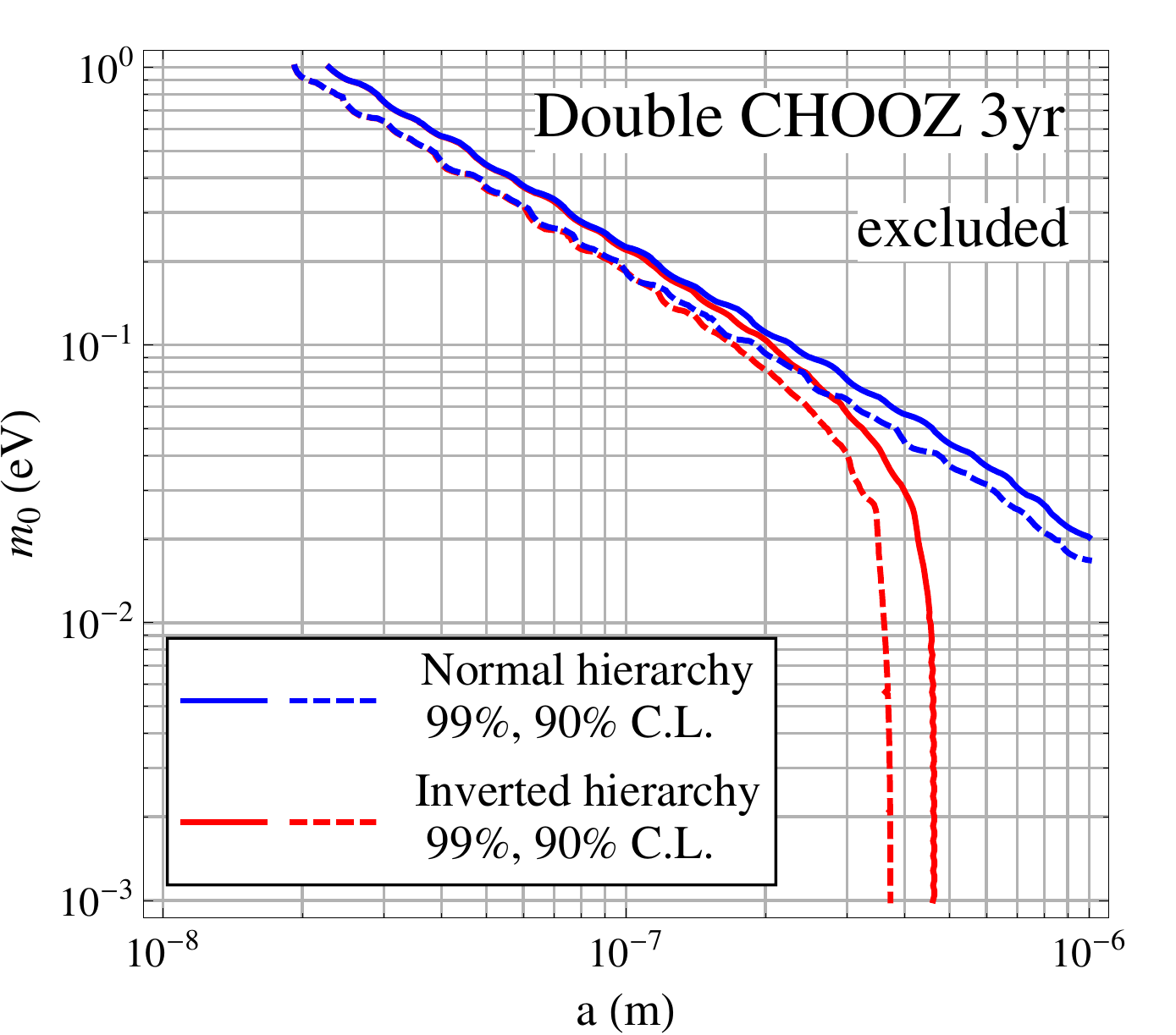}
\end{center}
\vglue -0.6cm
\caption{Sensitivity to LED predicted for Double CHOOZ after 3 years
  of data taking.}
\label{fig:dc-led}
\end{figure}

We also have estimated the improvement that this experiment can
provide on the limits given by CHOOZ and KamLAND. In
Fig.~\ref{fig:dc-led} we plot the potential exclusion region on the $a
- m_0$ plane As in the case of CHOOZ and KamLAND (see
Figs.~\ref{fig:chooz-led} and \ref{fig:kl-led}), we obtained the
better sensitivity for the IH case.
For some numerical limits, see Table~\ref{table:limits}.

\subsection{Accelerator $\nu_\mu \to \nu_\mu$ Experiments: T2K and NO$\nu$A}

\begin{figure}[bhtp]
\begin{center}
\includegraphics[width=0.48\textwidth]{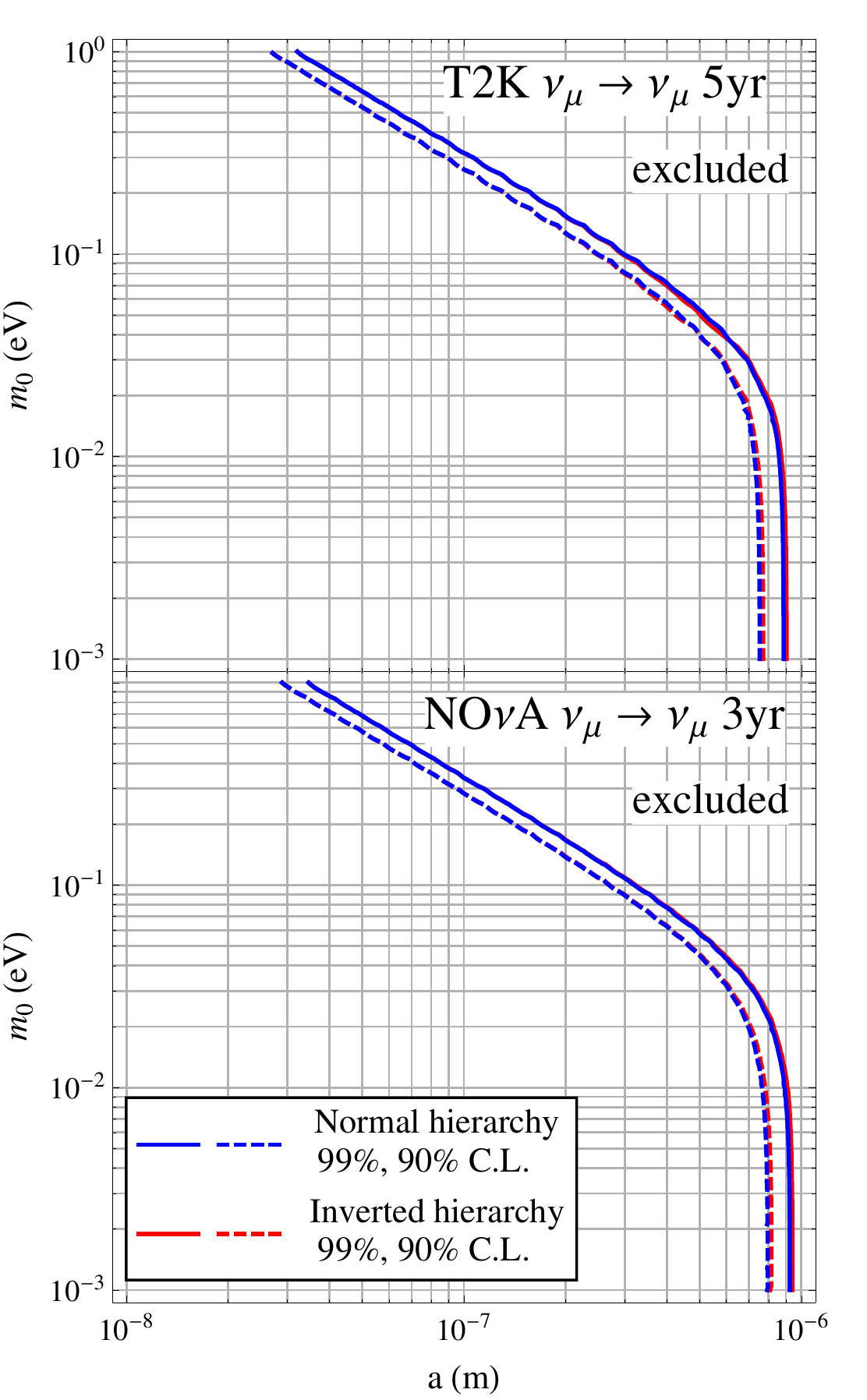}
\end{center}
\vglue -0.6cm
\caption{Sensitivity to LED predicted for T2K (top panel) and NO$\nu$A 
  (bottom panel) after 5 and 3 years of data, respectively.}
\label{fig:nova-t2k-led}
\end{figure}

T2K (Tokai to Kamioka)~\cite{T2K} is an experiment currently running
in Japan using a 0.75 MW $\nu_\mu$ beam from the J-PARC facility aimed
at the 22.5 kt water Cherenkov detector Super-Kamiokande with a 295 km
baseline. T2K in its first phase will take data in $\nu_\mu \to
\nu_{\mu,e}$ mode.

NO$\nu$A (NuMI Off-Axis $\nu_e$ Appearance)~\cite{Yang_2004}, is an
experiment that is currently being built in Fermilab and it will
observe $\nu_\mu \to \nu_{\mu,e}$ and $\bar\nu_\mu \to
\bar\nu_{\mu,e}$. The experiment will consist of a 222 t totally
active scintillator detector (TASD) near detector and a 25 kt TASD far
detector at 810 km and 1.12 MW of beam power.

We have simulated these experiments according to
Appendix~\ref{appendix-simulations}, considering 5 and 3 years of
$\nu_\mu \to \nu_\mu$ data for T2K and NO$\nu$A, respectively. In
fitting the data we have varied $\vert\Delta m^2_{31}\vert$ and
$\sin^22\theta_{23}$ freely, and considered priors on all other
standard parameters.

In Fig.~\ref{fig:nova-t2k-led} we show the potential excluded region
by T2K (5 yr) and NO$\nu$A (3 yr) in the $a - m_0$ plane The
limits are basically the same for those two experiments and they do
not depend on the mass hierarchy. For some numerical limits, see 
Table~\ref{table:limits}.

We see that neither of these experiments can really improve MINOS
limits. The reason for that is the fact that LED induces oscillations
into KK modes which are more sizable at higher energies away from the
oscillation minimum (see Fig.\ref{fig:prob}) as the probability is
larger in this region.  T2K and NO$\nu$A are narrow (off-axis) beam
experiments designed to measure precisely mixing parameters from the
behaviors of oscillation probabilities around the first oscillation
minimum, which means they are not very sensitive away from it.

\begin{table}
\begin{centering}
\begin{tabular}{|l|c|c|c|}
\hline 
 & \multicolumn{3}{c|}{Limit on $a$ ($\mu$m) at 90\% (99\%) C.\ L.\ }\tabularnewline
\hline 
Experiment & NH, $m_{0}\to0$ & IH, $m_{0}\to0$ & $m_{0}=0.2$ eV\tabularnewline
\hline 
CHOOZ & \dots & 0.54(0.61) & 0.13(0.14)\tabularnewline
\hline 
KamLAND & \dots & 0.79(0.91) & 0.19(0.22)\tabularnewline
\hline 
MINOS & 0.73(0.97) & 0.73(0.97) & 0.12(0.16)\tabularnewline
\hline 
Combined & 0.75(0.98)  & 0.49(0.57)  & 0.10(0.12) \tabularnewline
\hline 
Double CHOOZ & \dots & 0.38(0.46) & 0.09(0.11)\tabularnewline
\hline 
T2K & 0.76(0.89) & 0.76(0.89) & 0.13(0.16)\tabularnewline
\hline 
NO$\nu$A & 0.80(0.92) & 0.80(0.92) & 0.14(0.17)\tabularnewline
\hline
\end{tabular}
\par\end{centering}

\caption{Limits on the size $a$ of the extra dimension for both
  hierarchies and degenerate neutrinos. See text for more details.}
\label{table:limits}
\end{table}

\section{Discussion and Conclusions}
\label{sec:conclusions}

We have investigated the effect of LED in neutrino oscillation
experiments assuming that singlet SM fermion fields can propagate in
the bulk of a $d$-dimensional spacetime and couple to the SM neutrino
fields that lie in the brane through Yukawa couplings with the Higgs.
We have shown that terrestrial neutrino oscillation experiments can
provide submicrometer limits on the largest extra dimension $a$.

For hierarchical neutrinos with $m_0 \to 0$, CHOOZ, KamLAND and MINOS
together constrain $a < 0.75 (0.98)$ $\mu$m for NH and $a < 0.49
(0.57)$ $\mu$m at 90 (99)\% C.\ L.\  for IH.  For degenerate neutrinos with
$m_0= 0.2$ eV their combined data constrain $a < 0.10 (0.12)$ $\mu$m
at 90 (99)\% C.\ L.

We have also found that the future Double CHOOZ experiment will be 
able to improve these limits by roughly 20\% for the IH and 
10\% for the degenerate case. However, T2K and NO$\nu$A due to their 
narrow beam cannot surpass MINOS limits.

Let us discuss briefly what we can expect from solar and atmospheric
neutrinos.  For solar neutrinos, if the ratio $1/a$ is much larger
than $\sqrt{\Delta m^2_{21}}$, 
the matter effect is not important 
as long as the impact of LED on the standard oscillation is concerned, 
and the effect of LED is to induce vacuum like oscillations from active 
to sterile states, simply reducing the overall $\nu_e$ (or all active
$\nu$) survival probability. Since the inverse of the bound we
obtained from MINOS and KamLAND on $a$ is much larger than
$\sqrt{\Delta m^2_{21}}$, we expect only a small reduction of solar
$\nu_e$ due to LED which would not spoil the goodness of fit of solar
neutrinos by the standard MSW effect.  In fact in order to induce
strong distortion of the solar neutrino spectra, the size of $a$
should be in the range of $\sim (60-100)$ $\mu$m
\cite{Dvali:1999cn}, much larger than the bound we obtained.
Therefore, we expect that addition of the solar neutrino data to our
analysis would not improve much, if at all, the bound on the size of
the LED we obtained in this paper.
 
For atmospheric neutrinos, we have checked that for given values of
the size of the LED ($a$) and the lightest neutrino mass ($m_0$) and
mass hierarchy, the magnitude of the impact of LED on the $\nu_\mu \to
\nu_\mu$ ($\bar{\nu}_\mu \to \bar{\nu}_\mu$) and $\nu_e \to \nu_e$
($\bar{\nu}_e \to \bar{\nu}_e)$ survival probabilities are similar to
what we see in Fig.~\ref{fig:prob} for the relevant range of $L/E$
from $1-10^4$ km/GeV. This was done including earth matter effects
making use of the formalism presented in
Appendix~\ref{appendix-solution} . We have verified that, as long as
we consider parameters excluded by MINOS and/or KamLAND (shown in
Figs. 4-6), LED does not make the oscillation probability deviate
strongly from the standard oscillation scheme for atmospheric
neutrinos. Therefore, by adding atmospheric neutrino data to our
analysis, we do not expect significant improvement on the bounds
obtained on LED in our paper.

Let us try to make some comparison of our bounds with the ones from the LHC.
In LED models the connection between the fundamental scale of gravity,
$M_D$, the number of extra dimensions $d$, the size of the compactification
radius $a$ and the Planck mass $M_P$ is given by $M_D^{d+2}= M^2_P/(8\pi a^d)$
~\cite{PDG2010} where it was assumed that, for simplicity,
the size of the all LED radii $a$ is equal for $d \ge 2$.
So for $d=1$ ($d=2$) our limits on $a$ imply $M_D>10^6$ TeV ($M_D>22$ TeV).
On the other hand, the presence of LED also
predicts gravition-emission and graviton exchange
processes at colliders and according to Ref.~\cite{lhc-limits} ATLAS and
CMS at the LHC, after 36 pb$^{-1}$, can exclude $M_D< 3-4$ TeV, for $d=1$
and $2$, depending on the ratio $\Lambda/M_D$, $\Lambda$ being the
cutoff scale. Therefore, despite that the bounds we obtained in
this work are model dependent, so far, they are stronger than the ones that 
come from collider physics.

Recently, new flux calculations for reactor neutrinos became
available~\cite{Mueller:2011nm}. We took them into account in our
analysis of CHOOZ, KamLAND and Double CHOOZ but the impact 
of the change of the flux on our results is very small.
Nevertheless, with this new flux calculation, older reactor neutrino
oscillation experiments exhibit the so called {\em reactor antineutrino
anomaly} recently reported in Ref.~\cite{reactor-anomaly}.
We note that, although the inclusion in our analysis of these older
reactor neutrino oscillation experiments would not improve essentially
the limits obtained in this work, they could favor some range of the
LED parameters currently allowed (obtained in this work), and
therefore, deserve further study~\cite{MNZ}.

A final comment is in order. One cannot directly apply our limits to
models such as the one discussed in Ref.~\cite{Hallgren:2004mw}, where
neutrino oscillations are modified by the presence of reconstructed
nongravitational large extra dimensions. However, since the model
studied here is the continuum limit of the former, we suspect that
similar constraints could be derived in that case.

\vspace{-0.3cm}
\begin{acknowledgments} 
  \vspace{-0.3cm} This work is supported by Funda\c{c}\~ao de Amparo 
  \`a Pesquisa do Estado de S\~ao Paulo (FAPESP), Funda\c{c}\~ao de Amparo
  \`a Pesquisa do Estado do Rio de Janeiro (FAPERJ) and by Conselho
  Nacional de Ci\^encia e Tecnologia (CNPq).
The authors thank the theory group of Fermilab for their hospitality
during our visit to Fermilab under the summer visitor program of 2010
(HN and RZF) and the Program for Latin American Students 2010 (PANM)
where part of this work was done.
\end{acknowledgments}

\appendix 
\section{Solution of the Evolution Equation for a Constant Matter Potential}
\label{appendix-solution}

While matter effects are not very important for this work, in this
appendix, for the sake of completeness, we describe the solution of
the evolution equation in the presence of constant matter potential in
the context of large extra dimensions. See
also~\cite{Barbieri-et-al,Davoudiasl:2002fq} where a similar procedure
was adopted. We first diagonalize $m_{\alpha \beta}^{D}$ with respect
to the active flavors by defining the unitary transformations

\begin{eqnarray}
\nu_{\alpha L}^{\left(0\right)}\, & = & \, \sum_{i} U_{\alpha i} \, \nu_{i
  L}^{\left(0\right)},\qquad  
\nu_{\alpha R}^{\left(0\right)}\, =  \,
\sum_{i} R_{\alpha i}\, \nu_{i R}^{\left(0\right)}, \qquad \nonumber \\
\nu_{\alpha   R,\alpha L}^{\left(N\right)}\, &=& \, \sum_{i} R_{\alpha i}\, \nu_{i
  R,i L}^{\left(N\right)}, \; \; N \ge 1
\end{eqnarray}
so that $\sum_{\alpha \beta} U_{\alpha i}^{*} \, m_{\alpha\beta}^{D}\, R_{\beta j} =
  \delta_{ij} \,M_{i}$, with the lower case Roman indices $i,j
  =1,2,3$. Throughout this paper Greek indices will run over the 3 active
  flavors, Roman lower case indices over the 3 SM families and upper case 
Roman indices over the KK modes. Explicitly 
\begin{eqnarray}
a\, M_{i} & = & \lim_{N\rightarrow\infty}\left(\begin{array}{ccccc}
m_{i} a & 0 & 0 & \dots & 0\\
\sqrt{2}\, m_{i} a & 1 & 0 & \dots & 0\\
\sqrt{2}\, m_{i} a & 0 & 2 & \dots & 0\\
\vdots & \vdots & \vdots & \ddots & \vdots\\
\sqrt{2}\,m_{i} a & 0 & 0 & \dots & N \end{array}\right) \nonumber \\
&= & \lim_{N\rightarrow\infty}\left(\begin{array}{ccccc}
\frac{\sqrt{2}}{2} \xi_i & 0 & 0 & \dots & 0\\
\xi_i & 1 & 0 & \dots & 0\\
\xi_i & 0 & 2 & \dots & 0\\
\vdots & \vdots & \vdots & \ddots & \vdots\\
\xi_i & 0 & 0 & \dots & N \end{array}\right)
,
\label{eq:aM}
\end{eqnarray}
where $\xi_i = \sqrt{2} \, m_{i} \, a$.

Let us define the following states 
\begin{equation}
\tilde{\nu}_\alpha  \equiv \left(\nu_{\alpha}^{\left(0\right)} \, 
\nu_{\alpha}^{\left(1\right)} \, \nu_{\alpha}^{\left(2\right)} \dots \right)^{T}, \,
\quad \alpha=e,\mu,\tau
\end{equation}
\begin{equation}
\tilde{\nu}_i  \equiv \left(\nu_{i}^{\left(0\right)} \, 
\nu_{i}^{\left(1\right)} \, \nu_{i}^{\left(2\right)} \dots \right)^{T}, \,
\quad i=1,2,3 
\end{equation}
so that 
\begin{equation}
\left(\begin{array}{c}
\tilde{\nu}_{e}  \\
\tilde{\nu}_{\mu} \\
\tilde{\nu}_{\tau} 
\end{array}
\right)
= \mathcal{U}\; 
\left(\begin{array}{c}
\tilde{\nu}_{1}  \\
\tilde{\nu}_{2}  \\
\tilde{\nu}_{3} 
\end{array}
\right)
\, ,
\end{equation} 
where 
\begin{equation}
\mathcal{U} = \left(\begin{array}{cc|cc|cc}
U_{e1} & 0 & U_{e2} & 0 & U_{e3} & 0\\
0 & R_{e1} & 0 & R_{e2} & 0 & R_{e3}\\
\hline U_{\mu 1} & 0 & U_{\mu 2} & 0 & U_{\mu 3} & 0\\
0 & R_{\mu 1} & 0 & R_{\mu 2} & 0 & R_{\mu 3}\\
\hline U_{\tau 1} & 0 & U_{\tau 2} & 0 & U_{\tau 3} & 0\\
0 & R_{\tau 1} & 0 & R_{\tau 2} & 0 & R_{\tau 3}\end{array}\right)
\, .
\end{equation}

This allows us to write the neutrino evolution equation in matter as 
\begin{widetext}
\begin{equation}
i\frac{d}{dt}
\left(\begin{array}{c}
\tilde{\nu}_{1}  \\
\tilde{\nu}_{2}  \\
\tilde{\nu}_{3}  
\end{array}\right)_L = \left[
\frac{1}{2E}
\left(\begin{array}{ccc}
M_{1}^{\dagger}M_{1} & 0 & 0 \\
0 & M_{2}^{\dagger}M_{2} & 0\\
0 & 0 & M_{3}^{\dagger}M_{3}\end{array}\right)
+ \mathcal{U^\dagger}\,
\left(\begin{array}{ccc}
{\cal {V}}_e & 0 & 0 \\
0 & {\cal{V}}_\mu & 0\\
0 & 0 & {\cal {V}}_\tau \end{array}\right)\, 
\mathcal{U} \right]
\left(\begin{array}{c}
\tilde{\nu}_{1}  \\
\tilde{\nu}_{2}  \\
\tilde{\nu}_{3}  
\end{array}\right)_L
\, ,
\label{eq:evol}
\end{equation}
\end{widetext}

where $E$ is the neutrino energy and we have defined

\begin{equation}
{\cal{V}}_\alpha = 
\left(
\begin{array}{cc}
V_\alpha & 0 \\
0 & 0 
\end{array}
\right)
=
\left(
\begin{array}{cc}
\delta_{e \alpha}\,V_{\text{CC}}+V_{\text{NC}} & 0 \\
0 & 0 
\end{array}
\right) \, ,
\end{equation}
with the matter potentials $V_{\text{CC}}= \sqrt{2} \, G_F \, n_e$
and $V_{\text{NC}}= -\frac{\sqrt{2}}{2} \, G_F \, n_n$.  $G_F$ is the
Fermi constant, $n_e$ ($n_n$) is the electron (neutron) number density
in the medium and NC stands for neutral current.

Here we describe how to obtain an analytic expression for the
eigenvalues $\lambda_i^{(N)}$ and the amplitudes $W_{ij}^{(N0)}$
needed to calculate the transition amplitudes
${\cal{A}}(\nu_\alpha^{(0)}\to \nu_\beta^{(0)}; L)$ in
Eq.~(\ref{eq:amplitude}).

If we multiply Eq.~(\ref{eq:aM}) by its conjugate we get
\begin{widetext}
\begin{equation}
a^{2}M_{i}^{\dagger}M_{i}=\lim_{N\rightarrow\infty}\left(\begin{array}{ccccc}
\left(N+1/2\right)\xi_{i}^{2} & \xi_{i} & 2\xi_{i} & \dots & N\xi_{i}\\
\xi_{i} & 1 & 0 & \dots & 0\\
2\xi_{i} & 0 & 4 & \dots & 0\\
\vdots & \vdots & \vdots & \ddots & \vdots\\
N\xi_{i} & 0 & 0 & \dots & N^{2}\end{array}\right)=\left(\begin{array}{cc}
\eta_{i} & v_{i}\\
v_{i}^{T} & K\end{array}\right),
\end{equation}
\end{widetext}
where 
 \begin{equation}
\eta_{i}=\left(N+1/2\right)\xi_{i}^{2}\,,
\label{eq:etai}
\end{equation}
$v_{i}=\left(\xi_{i} \, 2\xi_{i} 
\, \dots N\xi_{i}\right)$ and $K=\mbox{diag}\left(1\;4\;9\;\dots N^2\right)$
 with $i=1,2,3$ the generation indices.  

Defining $V_{ij}= 2 E a^2 \displaystyle \sum_{\alpha=e,\mu,\tau} U_{\alpha i}^{*} U_{\alpha j} V_\alpha$
we can reorganize Eq.~(\ref{eq:evol})  as 
\begin{widetext}
\begin{equation}
\hskip -0.2cm
i\frac{d}{dt}\left(\begin{array}{c}
\nu_{1}^{\left(0\right)}\\
\nu_{2}^{\left(0\right)}\\
\nu_{3}^{\left(0\right)}\\
\nu_{1}^{\left(1\right)}\\
\nu_{2}^{\left(1\right)}\\
\nu_{3}^{\left(1\right)}\\
\nu_{1}^{\left(2\right)}\\
\nu_{2}^{\left(2\right)}\\
\nu_{3}^{\left(2\right)}\\
\vdots\\
\nu_{1}^{\left(N\right)}\\
\nu_{2}^{\left(N\right)}\\
\nu_{3}^{\left(N\right)}\end{array}\right)
\hskip -0.03cm 
=
\hskip -0.03cm 
\displaystyle \frac{1}{2Ea^2} 
\hskip -0.03cm 
\left(\begin{array}{ccc|ccc|ccc|c|ccc}
\eta_{1}+V_{11} & V_{12} & V_{13} & \xi_{1} & 0 & 0 & 2\xi_{1} & 0 & 0 & \dots & N\xi_{1} & 0 & 0\\
V_{21} & \eta_{2}+V_{22} & V_{23} & 0 & \xi_{2} & 0 & 0 & 2\xi_{2} & 0 & \dots & 0 & N\xi_{2} & 0\\
V_{31} & V_{32} & \eta_{3}+V_{33} & 0 & 0 & \xi_{3} & 0 & 0 & 2\xi_{3} & \dots & 0 & 0 & N\xi_{3}\\
\hline \xi_{1} & 0 & 0 & 1 & 0 & 0 & 0 & 0 & 0 & \dots & 0 & 0 & 0\\
0 & \xi_{2} & 0 & 0 & 1 & 0 & 0 & 0 & 0 & \dots & 0 & 0 & 0\\
0 & 0 & \xi_{3} & 0 & 0 & 1 & 0 & 0 & 0 & \dots & 0 & 0 & 0\\
\hline 2\xi_{1} & 0 & 0 & 0 & 0 & 0 & 4 & 0 & 0 & \dots & 0 & 0 & 0\\
0 & 2\xi_{2} & 0 & 0 & 0 & 0 & 0 & 4 & 0 & \dots & 0 & 0 & 0\\
0 & 0 & 2\xi_{3} & 0 & 0 & 0 & 0 & 0 & 4 & \dots & 0 & 0 & 0\\
\hline \vdots & \vdots & \vdots & \vdots & \vdots & \vdots & \vdots & \vdots & \vdots & \ddots & \vdots & \vdots & \vdots\\
\hline N\xi_{1} & 0 & 0 & 0 & 0 & 0 & 0 & 0 & 0 & \dots & N^{2} & 0 & 0\\
0 & N\xi_{2} & 0 & 0 & 0 & 0 & 0 & 0 & 0 & \dots & 0 & N^{2} & 0\\
0 & 0 & N\xi_{3} & 0 & 0 & 0 & 0 & 0 & 0 & \dots & 0 & 0 & N^{2}\end{array}\right)
\hskip -0.1cm
\left(\begin{array}{c}
\nu_{1}^{\left(0\right)}\\
\nu_{2}^{\left(0\right)}\\
\nu_{3}^{\left(0\right)}\\
\nu_{1}^{\left(1\right)}\\
\nu_{2}^{\left(1\right)}\\
\nu_{3}^{\left(1\right)}\\
\nu_{1}^{\left(2\right)}\\
\nu_{2}^{\left(2\right)}\\
\nu_{3}^{\left(2\right)}\\
\vdots\\
\nu_{1}^{\left(N\right)}\\
\nu_{2}^{\left(N\right)}\\
\nu_{3}^{\left(N\right)}\end{array}\right).
\label{eq:evolution}
\end{equation}
\end{widetext}

To diagonalize $\mathcal{H}$ we have to find the eigenvalues $\lambda_i^{(N)}$ 
that solve $\mbox{det}\left(2 E a^2\,\mathcal{H}-\lambda^{2}I\right)=0$. 
One can show, by using the Gauss algorithm for determinant calculation that 
this is equivalent to calculate 

\begin{equation} 
\mbox{det}\left(T\right)=0,
\label{eq:detT}
\end{equation}
where $T$ is a 3 by 3 matrix with elements

\begin{equation} 
T_{ij}=\left[-\lambda^{2}+\frac{\pi\xi_{i}^{2}\lambda}{2}\cot\left(\pi\lambda\right)\right]\delta_{ij}+ V_{ij}, (i,j=1,2,3).
\end{equation} 

To find the eigenvectors $w_{i}^{N}$, corresponding to the eigenvalues 
$\lambda_{i}^{(N)}$ we have to solve 
\begin{equation}
\mathcal{H}\,w_{i}^{N}=\lambda_{i}^{(N)2}\, w_{i}^N,
\label{eq:autoval}
\end{equation}
where we denote an element of $w_{i}^{N}$ by 
$(w_{i}^{N})^{M}_{j} \equiv W_{ij}^{(NM)}$. Explicitly in terms of these 
elements Eq.~(\ref{eq:autoval}) can be written as:
\begin{widetext}
\begin{equation}
\eta_{j}W_{ij}^{(N0)}+\sum_{A=1}^{K} A\, \xi_{j}W_{ij}^{(NA)}+\sum_{l=1}^{3}V_{jl}\, W_{il}^{(N0)}-(\lambda_i^{(N)})^{2}\, W_{ij}^{(N0)}=0,
\label{eq:wij0}
\end{equation}
\end{widetext}

and
\begin{equation}
A\, \xi_{j}W_{ij}^{(N0)}+\left(A^{2}-(\lambda_i^{(N)})^{2}\right)W_{ij}^{(NA)}=0.
\label{eq:wijA}
\end{equation}

We can obtain an equation for $W_{ij}^{(N0)}$ from Eqs.~(\ref{eq:wij0}),
and Eq.~(\ref{eq:wijA}) and Eq.~(\ref{eq:etai}) in the limit $N \to \infty$:

\begin{widetext}
\begin{equation}
W_{ij}^{(N0)}\left(\frac{\xi_{j}^{2}}{2}+\xi_{j}^{2}\sum_{A=1}^{\infty}
\frac{(\lambda_i^{(N)})^{2}}{(\lambda_i^{(N)})^{2}-A^{2}}-(\lambda_i^{(N)})^{2}\right)+\sum_{l=1}^{3}V_{jl}W_{il}^{(N0)}=0\end{equation}
\begin{equation}
\Leftrightarrow
W_{ij}^{(N0)}\left(\frac{\pi\xi_{j}^{2}\lambda_i^{(N)}}{2}\cot\left(\pi\lambda_i^{(N)}
\right)-(\lambda_i^{(N)})^{2}\right)+\sum_{l=1}^{3}V_{jl}w_{il}^{(N0)}=0\end{equation}
\end{widetext}

So that for each eigenvalue $\lambda_i^{(N)}$ obtained by solving 
Eq.~(\ref{eq:detT}) one has to solve
\begin{equation}
\sum_{l=1}^{3}T_{jl} W_{il}^{(N0)}=0,
\end{equation}
to obtain $W_{il}^{(N0)}$. We also need to impose the normalization of 
the eigenvector $w_{i}^{(N)}$ with 
\begin{widetext}
\begin{equation}
\sum_{l=1}^{3}\left\{ \left(W_{il}^{(N0)}\right)^{2}\left[1+\xi_{l}^{2}\left(\frac{\pi^{2}}{4}\cot^{2}\left(\pi\lambda_i^{(N)}\right)-\frac{\pi}{4\lambda_i^{(N)}}\cot\left(\pi\lambda_i^{(N)}\right)+\frac{\pi^{2}}{4}\right)\right]\right\} =1.
\end{equation}
\end{widetext}

As a technical note: in practice it is a very good approximation to
consider only the first five KK modes in the numerical calculation. We
have verified that the inclusion of higher modes do not cause any
significant change in our results.

In vacuum, $T_{ij}=T_i$ and $W_{ij}^{(N0)}=W_i^{(N0)}$ as the KK modes connected 
to different generations decouple. In this case, if $a^{-1}<< m_i$, as show in Ref.~\cite{Davoudiasl:2002fq}, we have

\begin{eqnarray}
\left( W_i^{(0N)}\right)^2 = \left \{ \begin{array}{cc}
 1-\frac{\pi^2}{6}\, \xi_i^2 + {\cal O}(\xi_i^4) & N=0 \\
\left(\frac{\xi_i}{N} \right)^2 + {\cal O}(\xi_i^4)  & N=1,2,3 ... 
\end{array}
\right .
\label{eq:vac}
\end{eqnarray}

\section{Simulation Details}
\label{appendix-simulations}

In this section we gather all information used to simulate the experiments.
We implemented all experiments using a modified version of GLoBES \cite{globes}.
To model the energy resolution, we used 
a following Gaussian smearing function, 
\begin{equation}
R\left(E,E^{\prime}\right)=\frac{1}{\sigma_{E}\sqrt{2\pi}}
e^{-\frac{\left(E-E^{\prime}\right)^{2}}{2\sigma_{E}^{2}}},
\end{equation}
where $\sigma_{E}$ was defined according to each experiment (see below).

Let us name the $\chi^2$ function without any uncertainty and
previous knowledge of oscillation parameters as $\chi^2_0$.  To
account for previous knowledge on some set of oscillation parameters
we use Gaussian priors.  Consider that these parameters $p_i$ have
mean values $\hat{p}_i$ and mean deviations $\sigma_{pi}$. Then, the
Gaussian priors are added to the $\chi^2$ as
\begin{equation}
\chi^{2} = 
\chi^{2}_0 + \sum_i\frac{\left(p_i-\hat{p}_i\right)^2}{\sigma_{pi}^{2}}.
\end{equation}

To deal with an experimental uncertainty (in flux, fiducial mass,
etc), we modify $ \chi^2_0 \rightarrow \hat{\chi}^2_0 $ by adding a
new parameter $x$ and add a penalty term $x^2 / \sigma_x^2$. To
exemplify that, let us assume an uncertainty $\sigma_{\rm NC}$ in the
neutral current events normalization.  If $N^{\rm NC}_i$ is the number
of neutral current events simulated at the $i$-th bin, then in the
$\chi^2_0$ function we could replace $N^{\rm NC}_i \rightarrow
(1+x_{\rm NC})N^{\rm NC}_i$ and add the penalty term $x_{\rm NC}^2 /
\sigma_{\rm NC}^2$ to the resulting $\chi^2$ function. In summary,
taking into account previous knowledge in the oscillation parameters
and experimental uncertainties, the resulting $\chi^2$ has the form

\begin{equation}
\chi^2 = \hat{\chi}^2_0 + \sum_i\frac{\left(p_i-\hat{p}_i\right)^2}{\sigma_{pi}^{2}} 
  + \sum_j \frac{x_j^{2}}{\sigma_{xj}^{2}}.
\end{equation}
For a detailed explanation about these techniques, 
see the GLoBES manual \cite{globes}.

Generically, for the data fits we have varied some of both standard
and LED oscillation parameters. When we considered Gaussian priors for
a standard parameter, we based the previous knowledge
on~\cite{concha}, using, at 1$\sigma$, $\Delta m_{21}^2 = 7.59 \pm
0.20 \times 10^{-5}$ eV$^2$, $\Delta m_{31}^2 = 2.46 \pm 0.12 \times
10^{-3}$ eV$^2$, $\theta_{12}=34.4^\circ \pm 1^\circ$,
$\theta_{23}=42.8^\circ \pm 4.7^\circ$ and we used a conservative
limit for $\theta_{13}$, $\sin^2 2\theta_{13}<0.09$. We did not impose 
any prior on $\delta_{CP}$.  For all fits with LED we varied freely $a$,
$m_0$ and the mass hierarchy.

It is useful to define the following quantities before giving the
details of each experiment. For KamLAND and MINOS, $N_i^{\rm exp}$ are
the experimental data points taken from Fig.1 of Ref.~\cite{kl11} and
Fig. 1 of Ref.~\cite{MINOS:Nu2010}, respectively, while for the future
experiments $N_{i}^{{\rm exp}}$ are the simulated data points
calculated assuming fixed values for oscillation parameters. Moreover,
$N_{i}^{\rm theo}$ are the theoretically calculated number of events
in the $i$-th energy bin which depend on the standard oscillation
parameters and, in the case of LED, also on $m_0$, $a$ and the
neutrino mass hierarchy. Given the complexity of the inclusion of
matter effects in the LED framework (see
Appendix~\ref{appendix-solution}), our simulations were done using the
vacuum oscillation probabilities. For KamLAND and MINOS this is
acceptable because the matter effects play a small role on the
survival channels. For CHOOZ and Double CHOOZ, since the baseline is
short, the matter effects are negligible. Finally for T2K and
NO$\nu$A, as long as we are fitting simulated data, the matter effects
are important only in the appearance channels, which are not used.

\subsection{CHOOZ}

In order to reproduce Fig. 55 of Ref.~\cite{Apollonio:2002gd}, we
considered a detector located at 1.05 km from the nuclear cores. The
predicted antineutrino spectrum was based on the newest fluxes
calculation available~\cite{Mueller:2011nm} and the overall
normalization was chosen so that the ratio between the observed and
theoretical unoscillated total number of events would match the value
given by~\cite{reactor-anomaly}, which is $R^{\rm exp}=0.961$.

Our analysis was based on rates information only, so we minimized a
$\chi^2$ function composed by

\begin{equation}
\chi^2_0 = \displaystyle 
\left( \frac{R^{{\rm exp}} - R^{\rm theo}}
     {\sigma} \right)^2,
\end{equation}
with respect to all parameters considered free in the fit. Here
$R^{\rm theo}$ is the ratio between the observed and theoretical
oscillated total number of events and $\sigma = 4.2\%$ takes into
account the statistical and systematical uncertainty.

\subsection{KamLAND}

We follow our previous papers~\cite{sado-ntz} in calculating the
number of events expected from reactors for a total exposure of 2881 t
yr. However, to calculate the unoscillated $\bar{\nu}_{e}$ spectrum we
have updated the averaged ratios of the fission yields of the four
isotopes that significantly contribute to the flux as: $^{235}$U:
$^{238}$U: $^{239}$Pu: $^{241}$Pu = 0.570: 0.078: 0.295: 0.057, in
accordance with Ref. \cite{kl11}. The energy resolution was modeled as a
Gaussian with $\sigma_{E}=0.064 \,\sqrt{ E/\mbox{MeV} - 0.8 }$.

We have determined the experimentally allowed regions minimizing 
the $\chi^2$ function composed by

\begin{equation}
\chi^2_0 = \sum_{i=1}^{17}{\displaystyle 
\frac{(N_{i}^{{\rm exp}}-N_{i}^{{\rm theo}})^2}{N_{i}^{{\rm exp}} +
  \sigma_{{\rm sys}}^{2}N_{i}^{\rm exp\,2}}},
\end{equation}
with respect to all parameters considered free in the fit. Here
$\sigma_{\rm sys} = 4.3\%$. The experimental data points were taken
from Ref.~\cite{kl11} in the energy window from 1.7 to 8.925 MeV (bin
width of 0.425 MeV). All uncertainty is included in 
$\sigma_{\rm sys}$ and we used the efficiency
given in \cite{kl11}.

\subsection{MINOS}

MINOS simulation was performed in accordance with \cite{Kopp:2010-qt},
using the NuMI neutrino beam given by \cite{Bishai:privcomm}, the
neutrino-nucleon cross section from
\cite{Messier:1999kj,Paschos:2001np}.  The analysis was performed with
neutrinos in 250 MeV bins from 1 to 5 GeV.  We assumed uncertainties
in the signal and background that were taken to be $\sigma_{\rm
  s}=4$\% and $\sigma_{\rm NC}=3$\%, respectively.  The detecting
efficiency was taken from Ref.~\cite{MINOS:Nu2010} and the energy
resolution was modeled as a Gaussian with $\sigma_{E}=0.16
\,E/\mbox{GeV}+0.07\, \sqrt{E/\mbox{GeV}}$ to best reproduce the MINOS
allowed region for the standard oscillation parameters.

We have determined the experimentally allowed regions minimizing the
$\chi^2$ function composed by
\begin{equation}
\chi^2_0=\sum_{i=1}^{16}{\displaystyle N_{i}^{{\rm
      exp}}\log\left(\frac{N_{i}^{{\rm exp}}}{N_{i}^{{\rm
        theo}}}\right)},
\end{equation}
with respect to all parameters considered free in the fit. 

\subsection{Double CHOOZ}

Basing Double CHOOZ simulation
on~\cite{Ardellier:2004ui,Huber:2009xx}, we used two identical 8.3 t
liquid scintillator detectors, one at 400 m and the at 1.05 km from
the nuclear cores.  The expected luminosity is 400 t GW y. We
considered 3 years of data taking assuming 62 energy bins from 1.8 to
8 MeV with the energy resolution modeled by a Gaussian with
$\sigma_{E}=0.12 \,\sqrt{ E/\mbox{MeV} - 0.8 }$. The uncertainties
taken into account for both cores and detectors were: isotopic
abundance (2\%), core power (2\%), flux normalization (0.6\%), overall
flux normalization (2.5\%) and energy scale for each core (0.5\%).

To estimate Double CHOOZ sensitivity we minimize the $\chi^2$ function
composed by
\begin{equation}
\chi^{2}_0=\sum_{i=1}^{62}\sum_{\rm d=N,F}\frac{\left(N_{\rm d,i}^{\rm
    exp}-N_{\rm d,i}^{\rm theo}\right)^{2}}{N_{\rm d,i}^{\rm exp} +
  \sigma_{\rm sys}^2 N_{\rm d,i}^{{\rm exp}2}},
\end{equation}
where $\sigma_{\rm sys}=1\%$, with respect to all parameters
considered free in the fit.  

\subsection{T2K}

We base T2K simulation on Ref.~\cite{Hiraide:2006vh} where we have
considered a beam power of 0.75 MW, a 22.5 kt water Cherenkov detector
at 295 km from the neutrino source, 5 years of data taking in the
$\nu_\mu \to \nu_\mu$ mode, 36 energy bins from 0.2 GeV to 2.0 GeV and
energy resolution modeled by a Gaussian with
$\sigma_{E}=80\,\mbox{MeV}$ for signal reconstruction and 2\%
uncertainty in the flux and background. For more details see
\cite{Hiraide:2006vh}.

To estimate T2K sensitivity we minimize the $\chi^2$ function
composed by
\begin{equation}
\chi^2_0 = \sum_{i=1}^{36}{\displaystyle 
\frac{(N_{i}^{{\rm exp}}-N_{i}^{{\rm theo}})^2}{N_{i}^{{\rm exp}}}},
\end{equation}
with respect to all parameters considered free in the fit. 

\subsection{NO$\nu$A}

 The experimental setup considered was based on
 \cite{Yang_2004,Huber:2009xx}, being a 25 kt TASD far detector at 810
 km, 1.12 MW of beam power, 3 years of data taking in the $\nu_\mu \to
 \nu_\mu$ mode, 20 energy bins from 1 GeV to 3.5 GeV and energy
 resolution modeled by a Gaussian with $\sigma_{E}=0.05
 \,\sqrt{E/\mbox{GeV}}$ for signal reconstruction and $\sigma_{E}=0.10
 \,\sqrt{E/\mbox{GeV}}$ for neutral current reconstruction.  We
 assumed uncertainties in the signal and background normalization
 using a slightly different method as discussed above. We used method
 ``C'' of GLoBES manual~\cite{globes} with $a$ and $b$ parameters
 (5\%:2.5\%) for both signal and background.

To estimate NO$\nu$A sensitivity we minimize the $\chi^2$ function
composed by
\begin{equation}
\chi^2_0 = \sum_{i=1}^{20}{\displaystyle 
\frac{(N_{i}^{{\rm exp}}-N_{i}^{{\rm theo}})^2}{N_{i}^{{\rm exp}}}},
\end{equation}
with respect to all parameters considered free in the fit.

\end{document}